\documentclass[11pt,a4paper,longauth]{article}

\usepackage[margin=2.5cm]{geometry}
\usepackage[T1]{fontenc}
\usepackage{graphicx}              
\usepackage{amsmath,amssymb}       
\usepackage{natbib}                
\usepackage{authblk}               
\usepackage[pdfencoding=auto,psdextra]{hyperref}
\hypersetup{
    colorlinks=true,
    linkcolor=blue,
    filecolor=magenta,      
    urlcolor=blue,
    citecolor=blue
}
\usepackage{graphicx}
\usepackage{txfonts}
\usepackage{xcolor}
\usepackage{lipsum}
\usepackage{subcaption}         
\usepackage{lscape}             
\usepackage{placeins}           
\usepackage[normalem]{ulem}     

\usepackage[markup=underlined]{changes}
\usepackage{xspace}

\usepackage{titling}

\usepackage{comment}

\newcommand{\FTS}{{FTS}\xspace}
\newcommand{\BBIR}{{BBIR}\xspace}
\newcommand{\BSM}{{BSM}\xspace}
\newcommand{\BSMs}{{BSMs}\xspace}

\newcommand{\HF}{{HF}\xspace}
\newcommand{\LF}{{LF}\xspace}

\newcommand{\FPA}{{FPA}\xspace}

\newcommand{\COBEF}{{\it COBE/FIRAS}\xspace}
\newcommand{\Planck}{{\it Planck}\xspace}
\newcommand{\ourmission}{{\it FOSSIL}\xspace}

\definecolor{fcorange}{HTML}{E37222}
\definecolor{fcorange}{HTML}{E37222}
\colorlet{changes}{fcorange}

\newcommand*\fnmsep{\unskip\hbox{\textsuperscript{\normalfont,}}}

\newcounter{inst}

\makeatletter
\newcommand{\inst}[1]{\unskip$^{#1}$}
\renewcommand{\and}{,\ } 

\newcommand{\institute}[1]{%
  \g@addto@macro\@author{%
    \\[0.8em] \normalfont\small
    \setcounter{inst}{1}%
    \renewcommand{\and}{\\[0.2em]\stepcounter{inst}$^{\theinst}$~}%
    $^{\theinst}$~#1%
  }%
}

\preauthor{\begin{flushleft}\large\raggedright}
\postauthor{\par\end{flushleft}}
\makeatother

\title{FOSSIL: A future mission for CMB spectral distortion measurements}

\author{
N. Aghanim\inst{1}\fnmsep\thanks{nabila.aghanim@cnrs.fr}
\and
B. Maffei\inst{1}
\and
J. Aumont\inst{2}
\and
A. Beelen\inst{3}
\and
B. Borgo\inst{1}
\and
E. Bozzo\inst{4}
\and
J. Chluba\inst{5}
\and
X. Coulon\inst{1}
\and
S. Couturier\inst{1}
\and
F. Cuttaia\inst{6}
\and
B. Cyr\inst{7}
\and
P. De Bernardis\inst{8,9}
\and
C. De Jabrun\inst{1}
\and
J.-J. D\'iaz-Garc\'ia\inst{10}
\and
D. Eckert\inst{4}
\and
G. Fabbian\inst{1}
\and
L. Ferrari\inst{11}
\and
F. Finelli\inst{6,12}
\and
T. Gascard\inst{13}
\and
J. E. Gudmundsson\inst{13}
\and
A. Kogut\inst{14}
\and
G. Lagache\inst{3}
\and
M. Loquet Le Gall\inst{1,15}
\and
S. Martin\inst{16}
\and
A. I. Silva Martins\inst{17}
\and
A. Monfardini\inst{18}
\and
C. O'Sullivan\inst{19}
\and
L. Pagano\inst{20,21}
\and
D. Paoletti\inst{6,12}
\and
G. Pisano\inst{8}
\and
M. Remazeilles\inst{22}
\and
V. Reveret\inst{23}
\and
J. A. Rubino-Martin\inst{10}
\and
V. Sauvage\inst{1}
\and
G. Savini\inst{24}
\and
L. Spencer\inst{25}
\and
A. Tartari\inst{26}
\and
L. Terenzi\inst{6}
\and
N. Trappe\inst{19}
\and
D. Watts\inst{17}
\and
B. Winter\inst{27}
\and
E. Baker\inst{28}
\and
E. S. Battistelli\inst{8,9,29}
\and
R. Battye\inst{5}
\and
J. L. Bernal\inst{22}
\and
A. Besnard\inst{1}
\and
F. R. Bouchet\inst{30}
\and
M. Braglia\inst{31}
\and
C. Burigana\inst{12,32}
\and
M. Calvo\inst{18}
\and
S. Caminade\inst{1}
\and
M. Citran\inst{33,34}
\and
D. Clements\inst{35}
\and
W. Coulton\inst{36}
\and
E. Courau\inst{15}
\and
G. De Zotti\inst{37}
\and
E. Di Valentino\inst{38}
\and
H. Dole\inst{1}
\and
G. Domenech\inst{39,40}
\and
S. Evangelista\inst{5}
\and
S. Farrens\inst{23}
\and
M. Frailis\inst{41}
\and
K. A. Glasscock\inst{17}
\and
V. Hervier\inst{1}
\and
E. Kovetz\inst{42}
\and
H. Liu\inst{28}
\and
T. Louis\inst{34}
\and
F. Madec\inst{3}
\and
A. S. Maniyar\inst{43}
\and
C. J. A. P.  Martins\inst{44,45}
\and
S. Masi\inst{8,9}
\and
L. Maurin\inst{1}
\and
F. Moll\inst{15}
\and
L. Morelli\inst{6,12,46}
\and
S. K. Naess\inst{17}
\and
A. Nicola\inst{5}
\and
F. Pace\inst{47,48,49,50}
\and
N. Ponthieu\inst{51}
\and
V. Poulin\inst{52}
\and
E. Pr\^ele\inst{18}
\and
N. Sanchez\inst{53}
\and
W. Qin\inst{54}
\and
L. Rodriguez\inst{23}
\and
D. Scott\inst{55}
\and
F. Simon\inst{56}
\and
J. Singal\inst{57}
\and
T. Slatyer\inst{7}
\and
E. M. Teixeira\inst{52}
\and
L. Thiele\inst{58,59}
\and
M. Tristram\inst{34}
\and
T. Trombetti\inst{12,32}
\and
L. Vacher\inst{34}
\and
S. Vinzl\inst{2}
\and
A. Zacchei\inst{41}
\and
E. Allys\inst{60}
\and
K. Aryan\inst{10}
\and 
J. M. Casas Gonzales\inst{61}
\and
O. Lahav\inst{24}
\and
C. Sidiropoulos\inst{62,63}}

\date{\textit{Author affiliations are listed in the Appendix.}}

\begin{document}

\maketitle

\begin{abstract}
\noindent
Measurements of Cosmic Microwave Background (CMB) spectral distortions represent a powerful and complementary window into the early Universe, dark matter physics, and the thermodynamic evolution of cosmic plasma and structures, probing physical processes far beyond the reach of the CMB temperature anisotropies. We present \ourmission, a space mission proposed for ESA's 8th Medium-class mission launch opportunity. \ourmission is designed to deliver unprecedented absolute spectroscopic measurements of the sky intensity over a broad frequency range (50\,GHz to 2\,THz) to accurately measure chemical-potential $\mu$-type, Compton $y$-type, and Compton relativistic $y$-type spectral distortions. 
The payload relies on a dual-input, dual-output differential polarising Martin-Puplett Fourier Transform Spectrometer (FTS) cooled to an operating temperature near 4.5\,K to minimise instrumental emissions. The focal plane utilises high-sensitivity, multi-moded Kinetic Inductance Detectors (KIDs) operated at 50\,mK. Absolute calibration and mitigation of optical and  thermal systematic effects are achieved through beam switching between the sky and a thermally regulated internal blackbody reference (BBIR) at a temperature close to that of the CMB. \ourmission will reach monopole sensitivities representing a gain of roughly three orders of magnitude over \COBEF. These capabilities will allow us to constrain primordial density fluctuations on scales $k \sim 10-10^4\text{~Mpc}^{-1}$, test inflationary models during the final 40 e-folds before the end of inflation, probe dark matter annihilation/decay/interactions, constrain primordial black hole seed scenarios, and pin down feedback models in galaxy evolution and structure formation by measuring the cosmic thermal energy content. Furthermore, its 130 spectral channels will enable legacy astrophysics, including precise characterisation of the Cosmic Infrared Background, line intensity mapping of [CII] and CO during cosmic noon, and detailed Galactic dust mapping.
\end{abstract}


\section{Introduction}
The Cosmic Microwave Background (CMB) is a cornerstone of cosmology, and measuring its anisotropies at scales ranging between a few degrees and an arcminute has been the focus of the community since the nineties. Over the last decades and throughout the past and present experiments, it has revolutionised our understanding of the Universe as it probes the origin of the initial density perturbations that give rise to the present-day structures and provides us with a standard model to describe the content of the Universe and its evolution.
Besides unveiling the nature of dark energy (DE), which is the purpose of the ongoing galaxy surveys DES\footnote{\url{https://www.darkenergysurvey.org/}}, DESI\footnote{\url{https://www.desi.lbl.gov/}}, \textit{Euclid}\footnote{\url{https://www.euclid-ec.org/}}, and LSST/Rubin\footnote{\url{https://rubinobservatory.org/}}, central challenges in cosmology lie in understanding the nature of dark matter (DM) and in probing the Universe at small scales to constrain the initial conditions, for example via B-mode polarisation in the CMB. Unlike structures at very large scales, which can be described by linear perturbation theory, small-scale phenomena involve nonlinear gravitational effects and complex processes associated with baryonic physics (e.g., star formation, feedback processes, and gas thermodynamics)  that are difficult to model. In addition, Silk damping erases primary anisotropies in the CMB at small scales, limiting the direct observational window into the primordial Universe. Finally, small scales also suffer from observational limitations that complicate theoretical predictions and data interpretation. As a result, key questions about the nature of DM, the physics of inflation, and the growth of structures remain unanswered. 

Probing the small-scale universe across different epochs, in both linear and non-linear regimes, requires new and independent observables. One of them is the measurement of deviations from the CMB  blackbody (BB) spectrum, namely the average (monopole) spectral distortions \citep{Chluba2019WPDEC,Voyage2050SDWP} and their anisotropies \citep{kite_spectro-spatial_2023-III}. These novel tracers of the early Universe, particle physics, and structure formation will allow us to probe otherwise inaccessible features, reaching beyond the epoch of recombination deep into the primordial universe \citep{Cyr2026SciencePaper}. The importance of this novel tracer was recognised as a possible target for future missions of the European Space Agency (ESA) as part of the Voyage 2050 programme\footnote{\url{https://www.cosmos.esa.int/web/voyage-2050}}, as one of the science priorities of the Centre National d'Etudes Spatiales (CNES) latest science survey\footnote{\url{https://cnes.fr/en/scientists/science-survey-seminar}}, and the ASTRONET 2022-2035 roadmap\footnote{\url{https://www.astronet-eu.org/wp-content/uploads/2023/05/Astronet_RoadMap2022-2035_Interactive.pdf}}. The increased interest in the measurement of spectral distortions highlights the vitality of the community, with several innovative concepts such as PIXIE, PRISTINE, and PRISM, proposed in the past decade \citep{pixie2011,PRISM2013WP,pixie2018,pristine2019}. In addition, three pathfinder experiments dedicated to CMB spectral distortion measurements are currently being developed: The Tenerife Microwave Spectrometer (TMS) \citep{TMS2020} is a ground-based spectrometer being developed to observe between 10 and 20\,GHz with commissioning at the Teide observatory planned for 2027. The COSmic Monopole Observer (COSMO) \citep{cosmo2021} is to be deployed in Antarctica, at the Dome-C Concordia base by 2029, where it will observe in two frequency bands covering 120--180\,GHz and 200--280\,GHz. Finally, the Balloon Interferometer for Spectral Observations of the Universe (BISOU), balloon-borne project covering frequencies between 90\,GHz and 1.5\,THz \citep{bisou2024}, is in Phase A at CNES with a planned first flight in 2030. These pathfinder projects will contribute to the increasing maturity of the instruments and measurement techniques while advancing scientific analysis and fostering synergies with the broader astronomy community.

However, only a space mission would be capable of fully excavating the exquisite data locked away in CMB spectral distortions. In this context, an ambitious mission is proposed thirty years after \COBEF  \citep{Fixsen:1996nj} to unambiguously measure CMB spectral distortions: \ourmission ({\it {F}TS f{O}r CMB {S}pectral di{S}tort{I}on exp{L}oration}). Building on the heritage of \COBEF, \ourmission is proposed to ESA in the context of the M8 call for a launch around 2041. It relies on measurements taken with a differential Fourier Transform Spectrometer (\FTS) that can provide an absolute measurement of the sky intensity with state-of-the-art kinetic inductance detectors. Proposed by a consortium of more than 270 researchers from 19 countries\footnote{Currently, France, Spain, Italy, Switzerland, UK, Netherlands, Germany, Norway, Ireland, Iceland, Canada, and the USA, in addition to ESA contribute to the design and/or procurement of payload elements.}, its core science objectives are articulated around the following primary goals: Probing early Universe physics and the inflationary paradigm, testing various black hole formation scenarios, understanding the nature of dark matter, and unveiling the hot gas content and its thermodynamics. Additional science goals can be enabled by the wide frequency coverage of \ourmission. They will improve our understanding of star formation in the first billion years of the history of the Universe, and of the properties of interstellar dust, providing definite discrimination between competing Galactic foreground models. 
The present paper provides an overview of \ourmission. In Section \ref{sec:science}, the main science goals are presented, followed by the description of the measurement technique (Sect. \ref{sec:measure}) and \ourmission payload (Sect. \ref{sec:PL}). Section \ref{sec:calib} describes the calibration strategy, Section \ref{sec:mission} gathers the main characteristics of the mission, and Section \ref{sec:data} outlines the data analysis principles needed for an instrument of this kind.

\section{Science case}\label{sec:science}
\begin{figure}[h]
\centering
\includegraphics[width=0.8\textwidth]{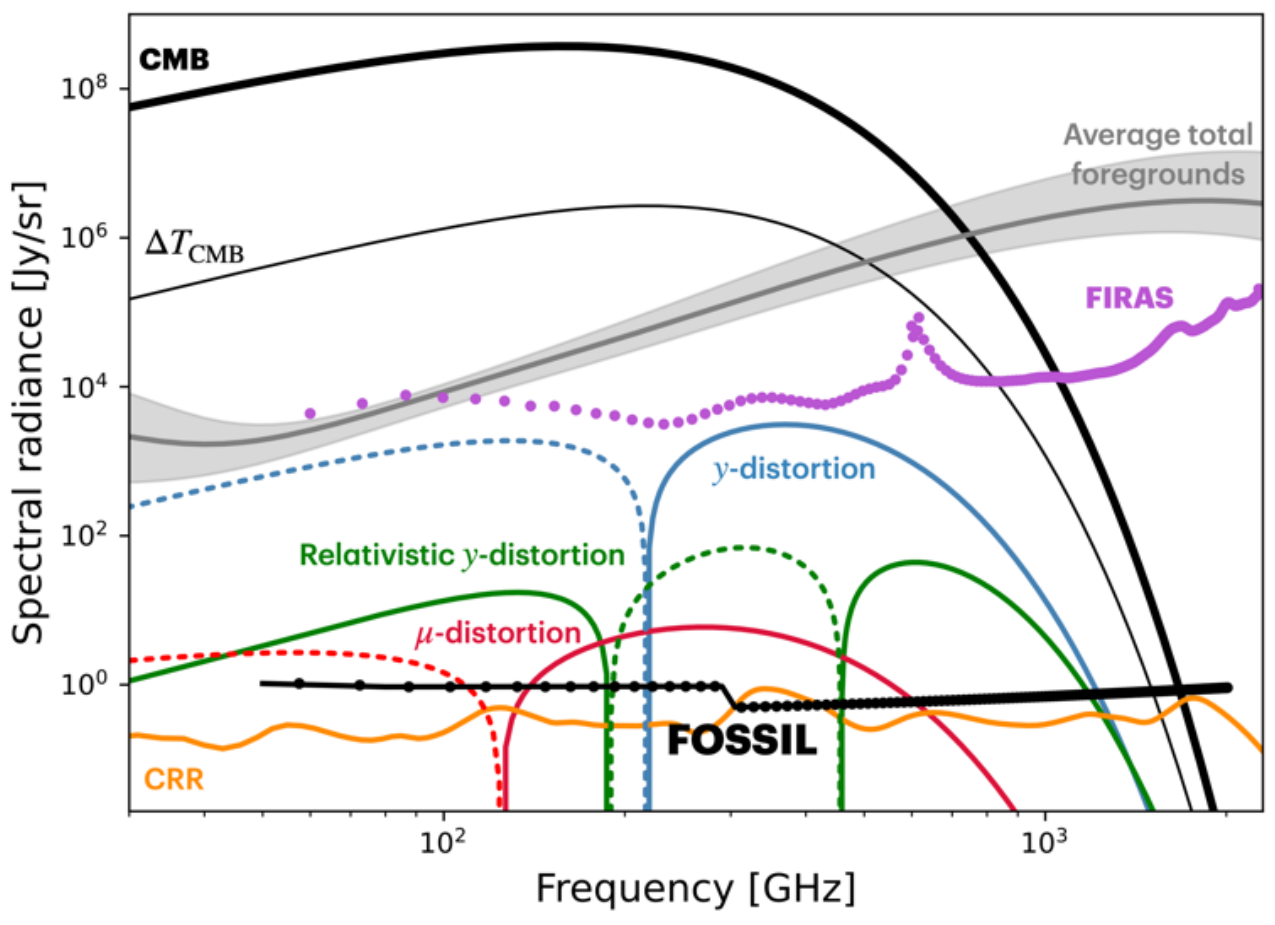}
    \caption{CMB spectral distortions (colored, solid curves correspond to photon excess relative to a blackbody, while dashed curves are photon decrements), cosmic recombination radiation, CRR (orange), and total foreground emissions (grey curve and area). \COBEF (purple) is compared to the \ourmission sensitivity (black). }
    \label{fig:SED} 
\end{figure}

Since \COBEF, the CMB spectrum is known to be a quasi-perfect BB. However, standard physics processes inevitably lead to spectral distortions that come in two main types: the chemical-potential $\mu$ distortions \citep{Sunyaev1970diss} imprinted at redshifts $z>5\times 10^4$  (Fig.~\ref{fig:SED}, red) and the Compton $y$ distortions \citep{Zeldovich1969,Sunyaev:1970er} originating from lower redshifts (Fig.~\ref{fig:SED}, blue), with rich thermalization physics interpolating between these extremes \citep{Illarionov1974, Danese1977, Danese1982, Burigana1991, Hu1993, Chluba2011, Khatri:2012tw, Chluba2013Green, Chluba2015GreensII, Acharya:2018iwh, Chluba2020large}.
In $\Lambda$CDM model, a $\mu$ distortion arises from the dissipation of the initial perturbations seeding galaxies and clusters. Its expected average (monopole) amplitude is predicted to be $\mu\simeq 2\times10^{-8}$ in the standard cosmological model  \citep[e.g.,][]{Chluba20122x2, Chluba2016}. The $y$-type spectral distortions arise from baryonic processes occurring during the reionisation of the Universe, interactions of CMB photons with hot gas in galaxy clusters, or for instance energy release from supermassive black holes. The expected average (monopole) amplitude is $y\simeq10^{-6}$ \citep{Refregier2000,daSilva2000, Hill_2015}. Additionally, relativistic corrections, which depend on the temperature of the electron gas $T_{\rm e}$ \citep{Wright1979, Rephaeli1995,Challinor1998, Sazonov1998,Itoh1998,Chluba2012SZpack}, are also visible in the average spectrum \citep{Hill_2015} (Fig.~\ref{fig:SED}, green). Finally, the process of cosmological recombination that occurred when the Universe was $\sim380,000$ years old leaves an unavoidable faint but characteristic pattern of spectral features (Fig.~\ref{fig:SED}, orange) due to the line emission from hydrogen and helium atoms \citep[e.g.,][for overview]{Sunyaev2009}.

Arising from different physical mechanisms at all scales and epochs, spectral distortions investigate both new physics and baryonic effects throughout the thermal history of the Universe in a unique manner \citep{Chluba2019WPDEC,Voyage2050SDWP,Cyr2026SciencePaper}. Building on the forecast measurement capabilities, and with a sensitivity around three orders of magnitude greater than \COBEF, we now present the main science goals that can be addressed with \ourmission. 

We summarise here the forecast analysis that was performed to estimate the detectability of the spectral distortion signals by \ourmission. It relies on a Fisher forecast approach that combines a sky model with an instrument model \citep{Coulon2024EPJWC,Coulon2024SPIE}. The former takes into account Galactic and extragalactic foreground contributions (Fig. \ref{fig:SED}, grey shaded area) modelled via the monopole emissions as a function of frequency. Based on \cite{abitbol_prospects_2017}, the detailed sky model used for the \ourmission forecast is described in \cite{Coulon2026-forecast}. The spectro-photometric model of the instrument is defined by relevant parameters describing the major subsystems \citep[see][]{Coulon2026-opti}. It is used to compute the associated photon-noise sensitivity shown as the black line in Fig. \ref{fig:SED}. 

The Fisher formalism estimates the uncertainties associated with the measurement of the CMB spectral distortions ($\mu$ and $y$) and all astrophysical parameters. It shows that the sensitivities $\sigma(y)\sim5\times10^{-9}$ and $\sigma(kT_e)\sim 8 \times10^{-3}\, \rm keV$ are easily within reach by \ourmission and that the $\mu$-distortion can be measured at a level $\sigma(\mu)\simeq1.5\times10^{-8}$ using only the monopole-frequency dependence (see \cite{Coulon2026-forecast} for details). This limit can be improved by the inclusion of spatial information. A recent re-analysis of \COBEF \citep{Bianchini:2022dqh,sabyr2025,Fabbian_2025} demonstrated that including spatial information to the monopole-only approach improves the sensitivity by a factor $\simeq$3--5 which would bring \ourmission's limit to $\sigma(\mu)\sim 5\times 10^{-9}$. The $\mu$-distortion signal at low frequencies lies where anomalous microwave emission (AME) \citep{Kogut1996, Oliveira1998, Lagache2003, Battistelli2019, Arce2020}, synchrotron \citep{Haslam1982,Bennett2003wmap_foregrounds,Planck2020components}, and free-free \citep{Dickinson2003ff_template, Draine2011Book} emissions dominate. Therefore, an improved extraction of $\mu$-distortion is further expected from synergies of \ourmission with low-frequency instruments. For instance, our forecasts \citep{Coulon2026-forecast} show that $15\%$ to $30\%$ improvement in the $\mu$-detection significance is expected when combining \ourmission with the 10--20\,GHz expected data from TMS.
\begin{figure*}[th]
\includegraphics[width=\textwidth]{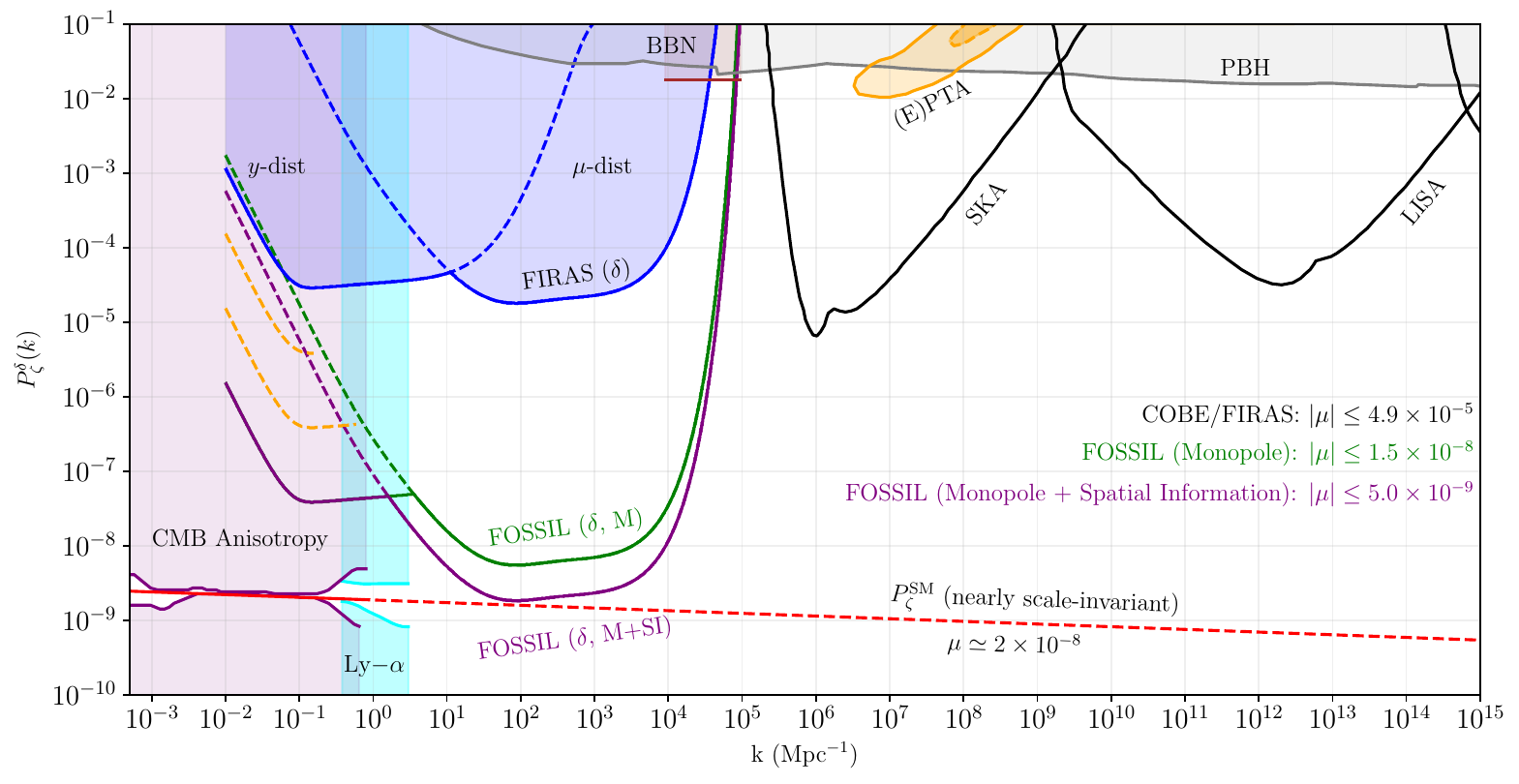}
    \caption{Primordial curvature power spectrum as a function of scale in $\Lambda$CDM and slow-roll single field inflation, with the dashed red line being the predicted power at small scales for a nearly scale-invariant initial spectrum. The forecasts, assuming a $\delta$-function spike at each $k$, for \ourmission from the monopole-only information are in green and the expected improvement by adding spatial information is in purple (details on the computation are found in ~\cite{Cyr:2023pgw}). Limits (coloured areas) come from CMB anisotropy~\citep{Planck:2018jri}, Lyman-$\alpha$ forest~\citep{Bird:2010mp}, \COBEF $\mu$- and $y$-type distortions~\citep{Chluba2012inflaton,
    Cyr:2023pgw, Pritchard:2025yda}, BBN \citep{Jeong2014} and PTA \citep{EPTA:2023fyk,EPTA:2023sfo}. The forecasted sensitivities of SKA and LISA are in black.
    }
    \label{fig:pspecs} 
\end{figure*}

\subsection{Testing the inflationary model and beyond}
The simplest inflationary model \citep{Starobinsky:1980te, Guth:1980zm, Linde:1981mu, Albrecht:1982wi} provides a widely accepted framework for generating primordial density perturbations seeding the formation of galaxies and clusters. Spectral distortions enable discrimination between different inflationary scenarios in a way that is independent and complementary to searches for CMB $B$-mode polarisation signals from primordial gravitational waves. 
For the standard nearly scale-invariant initial power spectrum, the dissipation of the initial perturbations through "Silk damping" implies a distortion of the CMB spectrum at a level $\mu\sim 2\times \,10^{-8}$ as shown in Fig.~\ref{fig:pspecs} \citep[e.g.,][]{Chluba20122x2,Chluba2012inflaton}. 
Detecting $\mu$ distortions with \ourmission will extend our grasp on primordial density perturbations on scales ranging from $k \sim 10-10^4\text{~Mpc}^{-1}$, more than three orders of magnitude smaller than those probed by standard CMB anisotropies and large-scale structure measurements (see Fig. \ref{fig:pspecs}). Our forecasted constraints for the $\mu$ signal in Fig. \ref{fig:pspecs} assumes a $\delta$-function spike at each wavenumber, $k$, in the primordial spectrum. The resulting forecast is thus very conservative. Given the integrated nature of the $\mu$-distortion, \ourmission will have the sensitivity required to observe the standard model spectrum, $\mu\sim 2\times \,10^{-8}$, as also shown in Fig. \ref{fig:SED} and in the Fisher forecast analysis \cite{Coulon2026-forecast}. Detecting the $\mu$ distortion will open an observational window 
probing modes that exited the horizon as late as roughly 40 e-folds before the end of inflation and providing us with a handle on parameters such as the running, or running of the running of the scalar spectral index \citep[e.g.,][]{Chluba20122x2, Khatri2013forecast, Chluba2013PCA, Chluba2013fore, Clesse2014, Cabass2016}. Alternatively, the measurement of the $\mu$-type distortion can test beyond-standard cosmological models and hypotheses such as primordial magnetic fields \citep[see for example,][] {Paoletti:2018uic,Jedamzik:2020krr,Thiele:2021okz,Galli:2021mxk,Paoletti:2022gsn} or cosmic strings \citep[e.g.,][]{Cyr2023}. The accurate measurements of the CMB energy density provided by \ourmission, when compared with that yielded by the standard primordial nucleosynthesis model \citep{Fields2020} will not only provide a consistency check of the model but can also allow us to derive unique constraints on the amount of energy that may have been dissipated in the redshift range from $z \simeq 3 \times 10^8$ (primordial nucleosynthesis) to $2-5 \times 10^6$ (minimum redshift from complete thermalization) \citep{DeZottiBonato2020}.\\
Additionally, the measurement of spectral distortions directly probes the thermal history of the Universe at redshifts before and during recombination via the characteristic pattern of spectral features across the microwave bands  \citep[e.g.,][]{Dubrovich1975, Dubrovich1997,RybickiDell94,Wong2006,Jose2006,Chluba2006b, Jose2008,Yacine2013RecSpec,Chluba2016CosmoSpec}.
\subsection{Nature of Dark Matter}
Identifying the nature of DM is one of the central challenges in modern cosmology and fundamental physics that \ourmission aims to tackle. Inferred from gravitational lensing and the CMB anisotropies, DM likely lies beyond the standard model of particle physics. A wide range of particle candidates have been proposed to account for DM, such as weakly interacting massive particles, axions and axion-like particles \cite[e.g.,][]{Marsh2016}), dark photons \cite[e.g.,][]{Fabbrichesi:2020wbt}, or more exotic alternatives \cite[e.g.,][]{Berlin:2022hmt,VanWaerbeke:2026sxs,Adams:2026xxx,Majidi2025}. Spectral distortions can provide a complementary test of these models through their distinct spectral footprints. The annihilation \citep[e.g.,][]{McDonald2000,Chluba2013fore}, decay  \cite[e.g.,][]{Chluba2013fore,Bolliet:2020ofj}, or scattering of DM particles with standard-model fields can heat and/or ionise the primordial plasma, modifying the CMB spectrum. \ourmission will aim to measure the associated $\mu$ and $y$ distortions and
hence probe DM properties such as the particle mass \cite[e.g.,][]{Ali-Haimoud:2015pwa,Ali-Haimoud:2021lka,Li:2024xlr}, lifetime \citep[e.g.,][]{Chluba2013PCA}, kinetic mixing parameters \cite[e.g.,][]{Chluba2024,Arsenadze2024, Cyr2024}, and interaction cross section \cite[e.g.,][]{Slatyer:2018aqg} at epochs inaccessible to laboratory experiments or large-scale-structure surveys.
\subsection{Origin of black holes}
Recent Pulsar-Timing-Array (PTA) results showed evidence for a cosmic background of gravitational waves (GWs) \cite[e.g.,][]{EPTA:2023sfo,InternationalPulsarTimingArray:2023mzf} that could be emitted by binary supermassive black holes (SMBHs). The latter are found at the centres of nearly all galaxies, but standard structure formation models lack an explanation for their origin. Several models propose that they began as massive ($\gtrsim 10^{3} \, M_{\odot}$) seed black holes. Primordial black holes (PBHs) \cite[see, e.g.,][for reviews]{Sasaki:2018dmp,Carr:2023tpt,Byrnes:2025tji}, which form from extremely large initial density perturbations, are natural candidates for these seeds. Other formation mechanisms are variants of the direct collapse scenario, in which molecular cooling in collapsing baryons is suppressed, leading to the formation of a large compact object \cite[e.g.,][]{Haiman:2012ic}, for example, when Population III star formation begins in halos with an additional source of radiation that destroys molecular hydrogen and the associated cooling. 
These scenarios can generate sizeable $\mu$ distortions \cite[e.g.,][]{Chluba2012inflaton,Chluba:2015bqa,Nakama:2017ohe,Byrnes:2018txb,taglia2025,Qin:2025ymc}. \ourmission can probe both primordial and baryonic seed-formation scenarios, with sizeable $\mu$ distortions providing a particularly sensitive signature of enhanced primordial small-scale power. 
In addition, it would complement the GW signal that will be measured in the LISA frequency range \citep{Bernal:2017nec}.

\subsection{Cosmic thermodynamics}
\begin{figure}[th]
\centering
\includegraphics[width=0.8\textwidth]{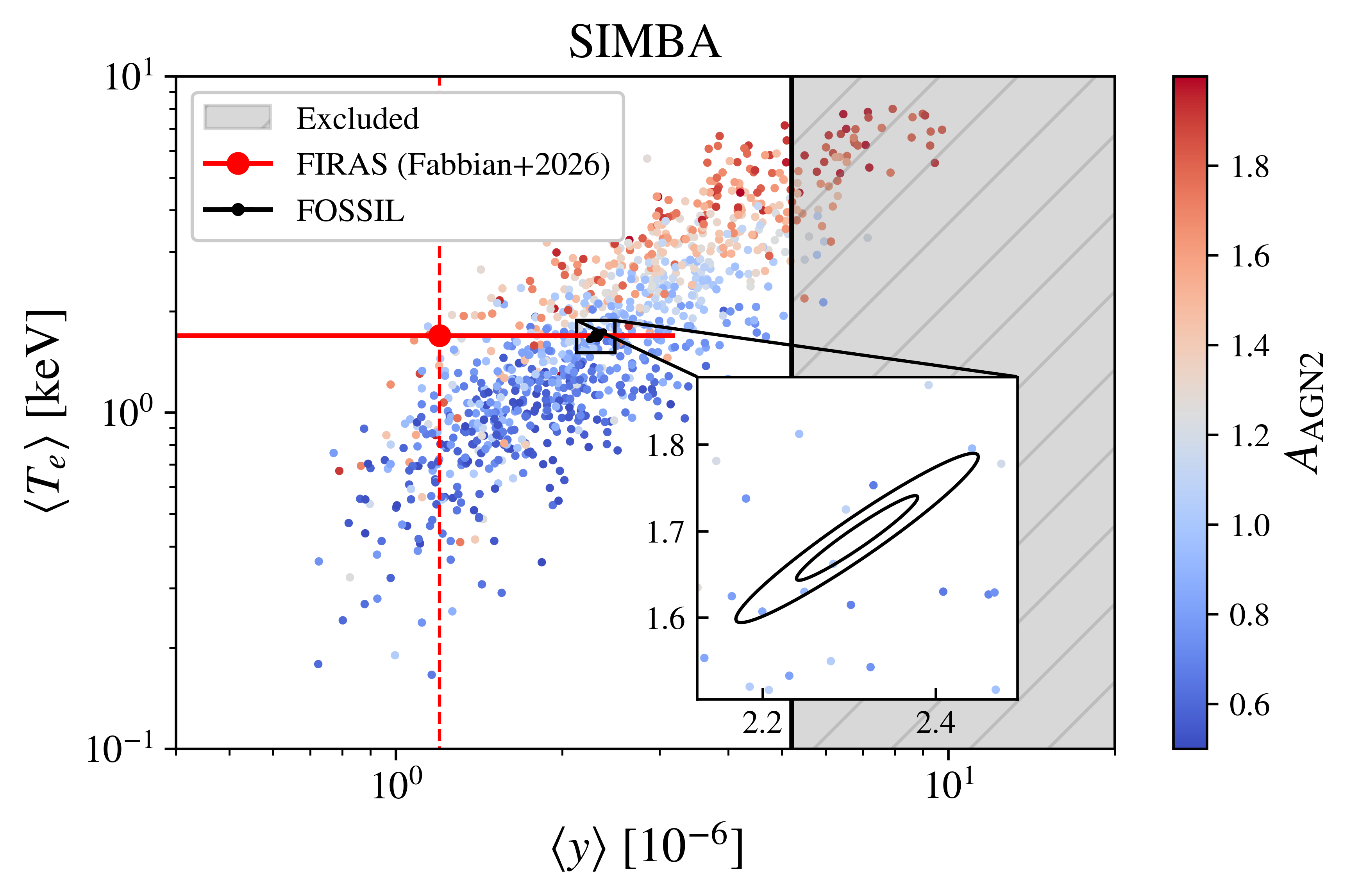}
    \caption{The forecast discrimination power of \ourmission of feedback models (black ellipse at 68 and 95\% C.L.). The points are predictions of $<T_e>$ and $<y>$ based on reprocessing of CAMELS simulations with the SIMBA feedback model \citep{2019MNRAS.486.2827D,2021ApJ...915...71V}. Each point shows results for a single simulation colour coded by the strength of the AGN feedback. The state-of-the-art measurements from \cite{Fabbian_2025} are shown in red together with the excluded area at 95\% C.L. in grey.}
    \label{fig:AGN2-ASN2} 
\end{figure}

Understanding large-scale structure (LSS) formation and evolution requires connecting the growth of DM halos to the complex baryonic physics governing galaxy and cluster formation. After decoupling, primordial density perturbations generated during inflation grew and evolved. Matter assembled into clusters and filaments, creating the cosmic web. The first stars and galaxies formed, eventually reionising the Universe. The conversion of baryons into stars, baryonic heating and cooling, and the spatial distribution of baryons is regulated by complex processes including gas cooling, metal enrichment, and multi-scale feedback from supernovae and active galactic nuclei (AGN). These feedback mechanisms redistribute energy and momentum, heating and displacing gas, suppressing star formation, and altering DM halo profiles, thereby coupling galaxy evolution to the thermodynamic state of large-scale environments.
Spectral distortions of the $y$-type are expected from inverse Compton scattering  \citep[also known as thermal Sunyaev-Zeldovich (tSZ) effect][]{Sunyaev_1972} of CMB photons with the free electrons of the ionised warm/hot gas during reionisation and inside the LSS \citep[e.g.,][]{Tanimura:2020lgs,Tanimura:2022uxh}. This guaranteed signal for \ourmission traces the line-of-sight integrated electron pressure of the ionised gas. Together with the relativistic tSZ signal \citep[e.g.,][]{Challinor1998,Itoh1998,Nozawa_1998, Chluba2012SZpack}, tracing the electron temperature $T_{\rm e}$, the tSZ effect enables a census of hot baryons and uniquely probes the thermodynamic state of ionised gas across cosmic time \citep{Hill_2015, Chiang2020y}. Following the approach by \citep{Thiele_2022,Fabbian_2025}, our forecast analysis (see Fig. \ref{fig:AGN2-ASN2}) shows that \ourmission could put sub-percent level constraints on the total integrated energy injected into the gas through the radiative and mechanical energy released from supernovae and SMBH-powered AGN, isolating or disproving proposed feedback models otherwise poorly constrained \citep[e.g.,][]{Thiele_2022, Valentini_2025}. The correlation with external tracers of the LSS at different redshifts could constrain the LSS formation process.
\ourmission's degree resolution corresponds to the size of the largest ionised bubbles at $z\sim7$ \citep[physical sizes of tens of Mpc, see][]{Neyer2024}. By the time \ourmission will operate, we expect that deep and wide area NIR multiband imaging and spectroscopic surveys will be available (first steps in this direction are the \textit{Euclid} Deep survey over an area of 50 sq. degrees and the ultradeep tier of the MOONRISE spectroscopic survey \citep{Maiolino2020}). A cross-correlation of \ourmission maps with the brightest galaxies at $z\sim6-7$, expected to be the signposts of over-densities associated with the ionised bubbles, will shed further light on the re-ionisation process. Furthermore, combining \ourmission with low-frequency data will also make it possible to improve our comprehension of the free-free cosmological distortion, which is predicted to occur both locally, originating from galaxy groups and clusters \citep{2011MNRAS.410.2353P}, and globally, generated by ionised matter mainly at the onset of the Universe's reionisation, when the efficiency of bremsstrahlung in a nearly uniform medium was still significant, and at low redshift, due to the amplification of bremsstrahlung caused by increased matter clumping \citep{2014MNRAS.437.2507T}.

\subsection{Primordial spectral distortions beyond the monopole}
While the spectral distortion science case focuses on the information that can be extracted from the average spectrum, much could be learned from  spectral distortion anisotropies \citep[e.g.,][]{kite_spectro-spatial_2023-III}. The latter are imprinted by two primary mechanisms: i) anisotropic sourcing of distortions and ii) the propagation of the (average) distorted CMB spectrum through a perturbed medium \citep{Chluba20122x2, Chluba:2022efq, kite_spectro-spatial_2023-III}. The detailed framework for computing these types of signals has only recently become available \citep{Chluba:2022xsd, Chluba:2022efq, kite_spectro-spatial_2023-III, Chluba2026}, allowing us to predict the auto- and cross-power spectra for various scenarios. A detection of the distortion auto-power spectra is beyond the current reach of experiments, but by measuring the distortion cross-power spectra with CMB temperature and polarisation anisotropies one could establish new probes of inflation \citep[e.g.,][]{Pajer2012b, Ganc:2012ae, Biagetti:2013sr, Emami:2015xqa, Ota:2016mqd, Dimastrogiovanni2016, Chluba2017, Ravenni:2017lgw, Chluba2026muT} and particle physics scenarios \citep[e.g.,][]{kite_spectro-spatial_2023-III, Evangelista2026}.

Much work is still needed on this new frontier, but the synergies with past, ongoing, and future CMB imaging experiments can be used to extract anisotropic distortions and place interesting constraints on the model space \citep{Remazeilles2018:mu, Rotti2022, Bianchini:2022dqh, kite_spectro-spatial_2023-III, Zegeye2023}.
\ourmission will advance the new frontier in two important ways: i) it will provide $\sim$degree-resolution maps of the distortion anisotropies benefitting from its wide frequency coverage and control of systematics to mitigate foregrounds and ii) by directly measuring the monopole distortions it will shed light on the origin of the distortion anisotropy that only an absolutely calibrated CMB spectrometer can achieve.\\
\ourmission's absolutely calibrated mapping of the sky on very large scale will enable us to perform multi-frequency tests regarding the origin of the low multipole patterns by exploiting their leakage \citep{2017PhRvL.119v1102Y, 2021A&A...646A..75T}, a topic linked to the lack of power at low multipoles, particularly in the quadrupole, first detected by COBE and confirmed by WMAP and {\it Planck}. Reconstructing the anisotropy power at very low multipoles is of fundamental importance for inflationary models that predict large-scale power suppression \citep[(see e.g.,][]{1982PhRvD..26.1231V, 1992JETPL..55..489S, 2003MNRAS.342L..72B, 2003MNRAS.343L..95E, 2006PhRvD..74d3518S} in relation to the Universe’s geometry and topology (see e.g., \cite{2002PhRvD..65d3513G,2002GReGr..34.1461E}).
\subsection{Star-formation history}
A substantial fraction of the star formation during the epoch known as “cosmic noon” ($1\leq z\leq4$) occurs within heavily dust-enshrouded galaxies \citep{MadauDickinson2014} and translates into the Cosmic Infrared Background (CIB) \citep{Puget1996}, the cumulative emission from dust-obscured star formation integrated over cosmic time. \ourmission frequency coverage, from 50\,GHz to 2\,THz, and sensitivity can provide a direct absolute measurement of the mean CIB intensity, improving upon two decades of efforts by, e.g, \cite{fixsen1998,lagache1999,bethermin2012,zavala2017,lagache2000,auclair2024,Fujimoto2024,gao2024}. The sub-percent precision on both the CIB monopole amplitude and spectral index, and the mK-level constraints on the effective dust temperature can quantify the contribution of obscured star formation in the first billion years of cosmic star-formation history.

Moreover, \ourmission's spectral resolution, $\Delta\nu=$15\,GHz will allow us to measure the integrated signal from redshifted CO spectral ladder and [CII] line emissions from galaxies during the cosmic noon \citep[$z\sim$ 1--4, see][]{Serra2016, 2017ApJ...838...82S, Chung2024}. The line-intensity mapping (LIM) on large angular scales with absolute calibration should provide a complete, volume averaged census, linking gas content to star formation history in a way that individual galaxies cannot \citep{Kovetz2017, Bernal2022, Chang2026}. In combination with LIM experiments, \ourmission will therefore provide key new insights for breaking existing modelling degeneracies \citep{Chung2024}.

\subsection{Galactic science}
\begin{figure}[th]
\centering
\includegraphics[width=0.8\textwidth]{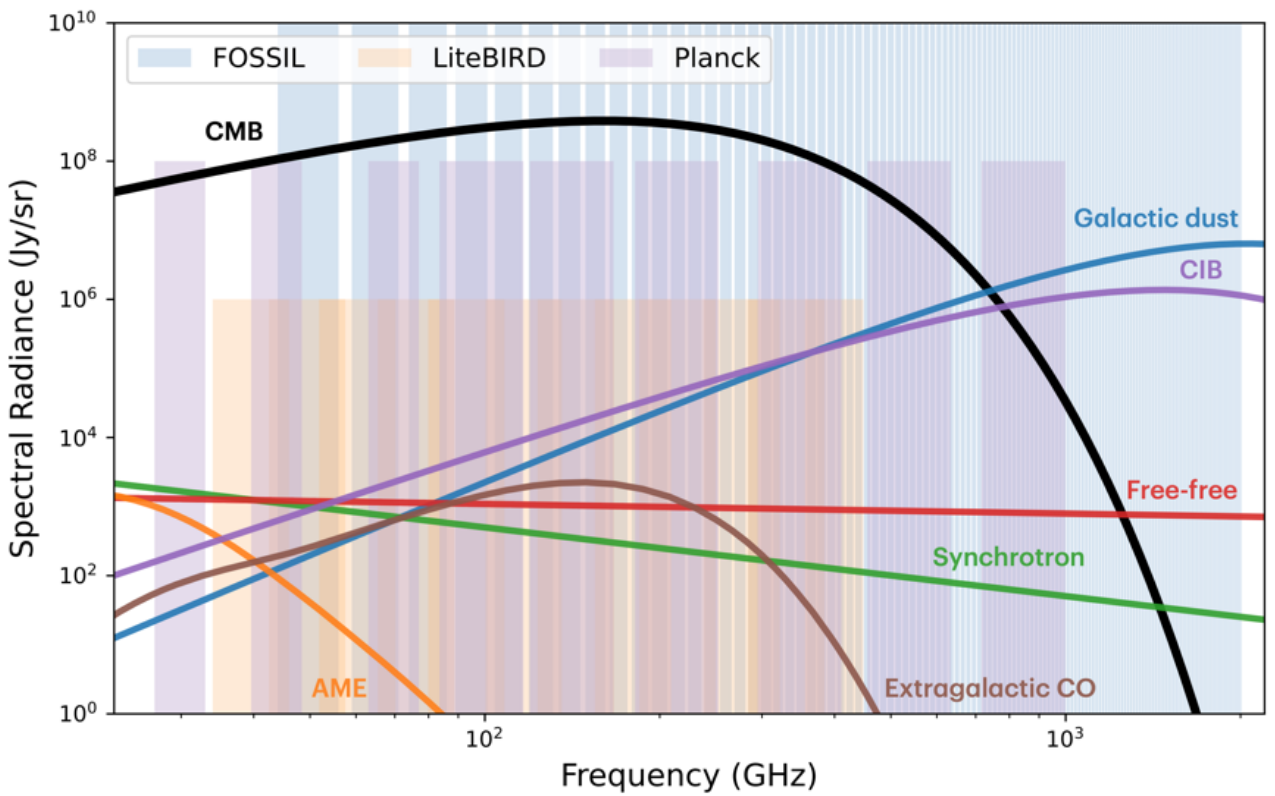}
    \caption{\ourmission's 130 channels compared with \textit{Planck} and LiteBIRD.
    Lines in red, green, orange, and blue are for free-free, synchrotron, AME, and Galactic dust. The brown and purple lines are for the extragalactic contributions, namely CO and CIB. The different error-bar heights are arbitrary and the gaps in frequency for \ourmission are for visual clarity only. }
    \label{fig:bands} 
\end{figure}

A detailed characterisation of the spectral energy distribution (SED) of the diffuse radio synchrotron background and the galactic emissions should be possible thanks to the highly sensitive absolute spectrometric survey of the full sky with 130 channels between 50\,GHz and 2\,THz that will be delivered by \ourmission  (Fig.~\ref{fig:bands}). 
\\
The unprecedentedly accurate measurement, at high frequencies ($\gtrsim 100$ GHz), of the thermal dust SED in intensity will yield critical insights on the microphysical properties of interstellar dust grains together with the lifecycle of matter, and the dynamical state of the ISM across Galactic environments. In particular, the measurement of the effective dust temperature with mK-level precision, and that of the dust amplitude and spectral index to sub-percent precision, will enable us to identify the proposed deviations from the single modified blackbody model \cite[e.g.,][]{Demyk2017,Guillet2018,Draine2021,Hensley2023}. In addition, bright emission lines (e.g., CO transition ladder, \textsc{C\,i}, \textsc{C\,ii}, \textsc{N\,ii}, \textsc{O\,i}, CH, H$_2$O, OH, etc.) will be mapped, probing the temperature, local density, and photoionisation rate in the Milky Way \citep{tielens1985}.
%
\section{Measurement principle}\label{sec:measure}
The measurement of CMB spectral distortions relies on an accurate absolute determination of the sky intensity as a function of frequency. The associated requirements are thus an absolute measurement, a broad frequency coverage (to capture the full spectral signature of distortions and astrophysical contributions), a coarse spectral resolution, and a modest to low angular resolution given that the monopole CMB spectral distortions are the primary science goals. The corresponding payload is described in Sect. \ref{sec:PL}. 

\begin{figure*}[!ht]
    \centering
    \includegraphics[width=\textwidth]{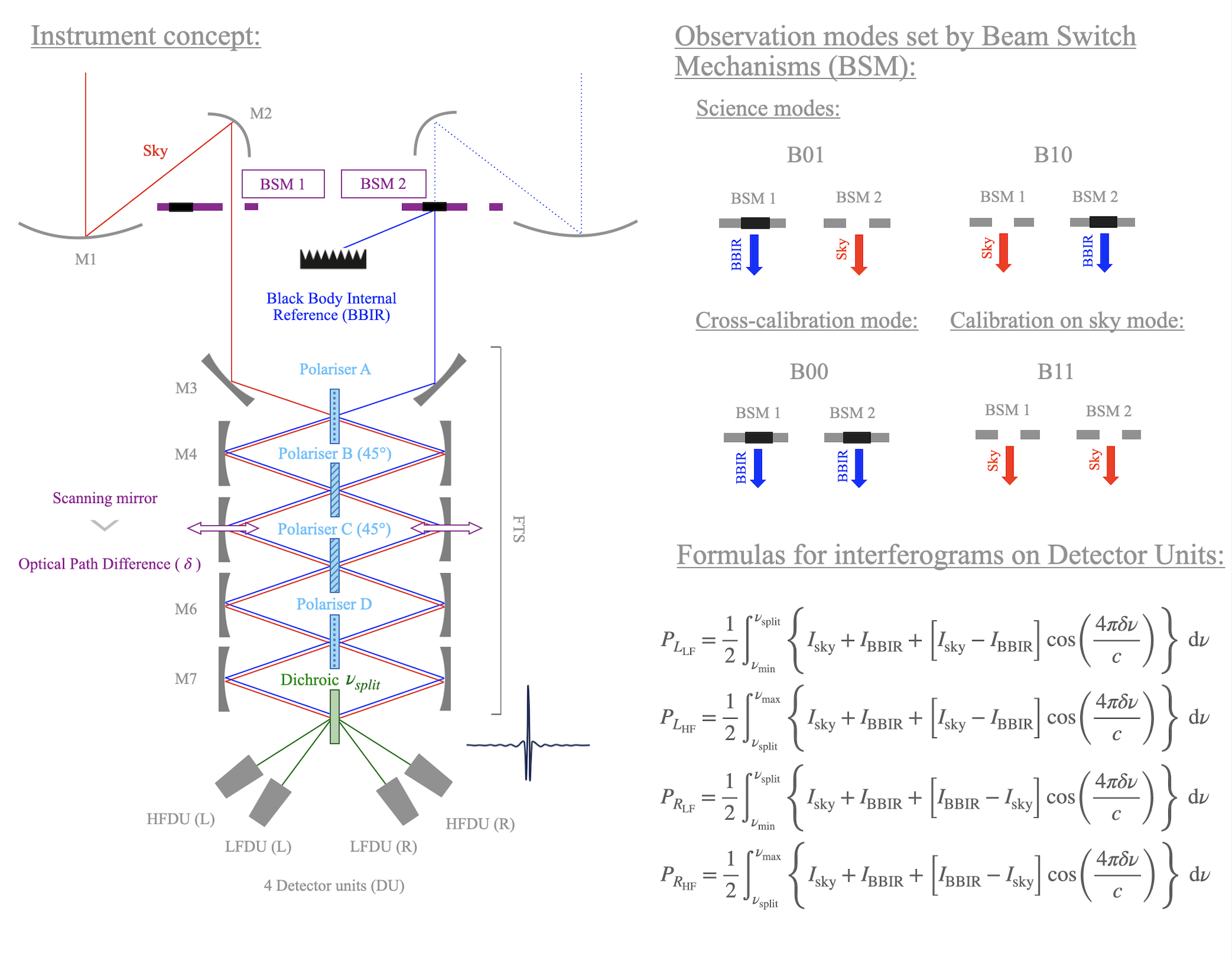}
    \caption{\textit{Left}: Conceptual view of the measurement method. The position and angle to the \BBIR are not representative of the actual intended optical coupling angle.
    \textit{Top right}: Observation modes set by the positions of the two beam switch mechanism (\BSM). 
    \textit{Bottom right}: Mathematical representation of the IFGs collected on each of the four detector units. They are representative of the ideal expressions for either B01 or B10 (with a change of +/- signs in the [] for Left/Right).    
    Here $\nu_{\rm min}$= 50\,GHz and $\nu_{\rm max}$= 2\,THz. $\nu_{\rm split}$ defined by a dichroic filter is set to 300\,GHz.}
    \label{fig:concept}
\end{figure*}

The concept of the measurement method (Fig.~\ref{fig:concept}) is inspired by that of the PIXIE space mission proposal \citep{Kogut2016SPIE}. \ourmission will map the sky through a continuous scanning strategy, spinning around its axis while performing absolute spectroscopic measurements with respect to a blackbody internal reference (\BBIR) across a frequency range of 50 - 2000\,GHz using a Fourier transform spectrometer (\FTS) with two inputs and two outputs. 

An interferogram (IFG) is acquired thanks to an \FTS scanning mirror mechanism (FTSM) introducing a variable optical path difference (OPD), $\delta$, between its two arms. Each FTS output is split into two spectral sub-bands by a dichroic filter. Multimode detectors collect these outputs presenting  a signal consisting of a constant and an FTS-scan-modulated part proportional to
the difference between the two input ports  (Fig.~\ref{fig:concept}, right). Fourier transformation of the IFG then give access to the spectrum of the difference between the two inputs. 
The ideal operation temperature for the entire instrument needs to be close to the CMB temperature, $T_{\rm CMB}=2.72548 \pm 0.00057$\,K \citep{Fixsen2009}, so that instrumental emission becomes negligible. The concept underlying \ourmission involves cooling the internal reference to the CMB temperature while the rest of the instrument is cooled to 4.5\,K—a temperature sufficiently close to $T_{\rm CMB}$ to minimise instrumental emission, thus allowing the removal of systematic effects through appropriate control and monitoring of the instrument's temperature and allowing the primary objectives to be met (see Sect. \ref{fig:BBref}). 
In addition, \ourmission takes full advantage of a few multi-moded feedhorn-coupled detectors instead of the thousands of detectors required for CMB imagers. The detector units are cooled at sub-Kelvin temperature.
The ultimate sensitivity requires strict control of the instrument systematic effects that may occur as a result of optical or thermal imbalances between the two arms of the \FTS. This can be achieved by selecting each of the two input ports of the \FTS to be directed either to the sky or the \BBIR such that four operation modes are performed (Fig.~\ref{fig:concept}, right): two modes are devoted to monitoring of systematics and the other two modes are dedicated to scientific observation. In the latter case, one of the \FTS input is directed towards the sky while the second input is directed towards the \BBIR without going through a telescope. With a perfectly symmetric instrument, the two output signals should be exactly the same. In the other two modes, either both \FTS inputs are directed towards the \BBIR and the acquired spectrum gives information about the differential response between the two paths after the telescope mirrors, or both of the \FTS inputs are looking at the sky through their respective telescopes and the acquired spectrum measures mismatches between the two paths, including the telescopes.
%
\section{FOSSIL Payload}\label{sec:PL}
\begin{figure*}[h]
    \centering
    \includegraphics[width=0.43\textwidth]{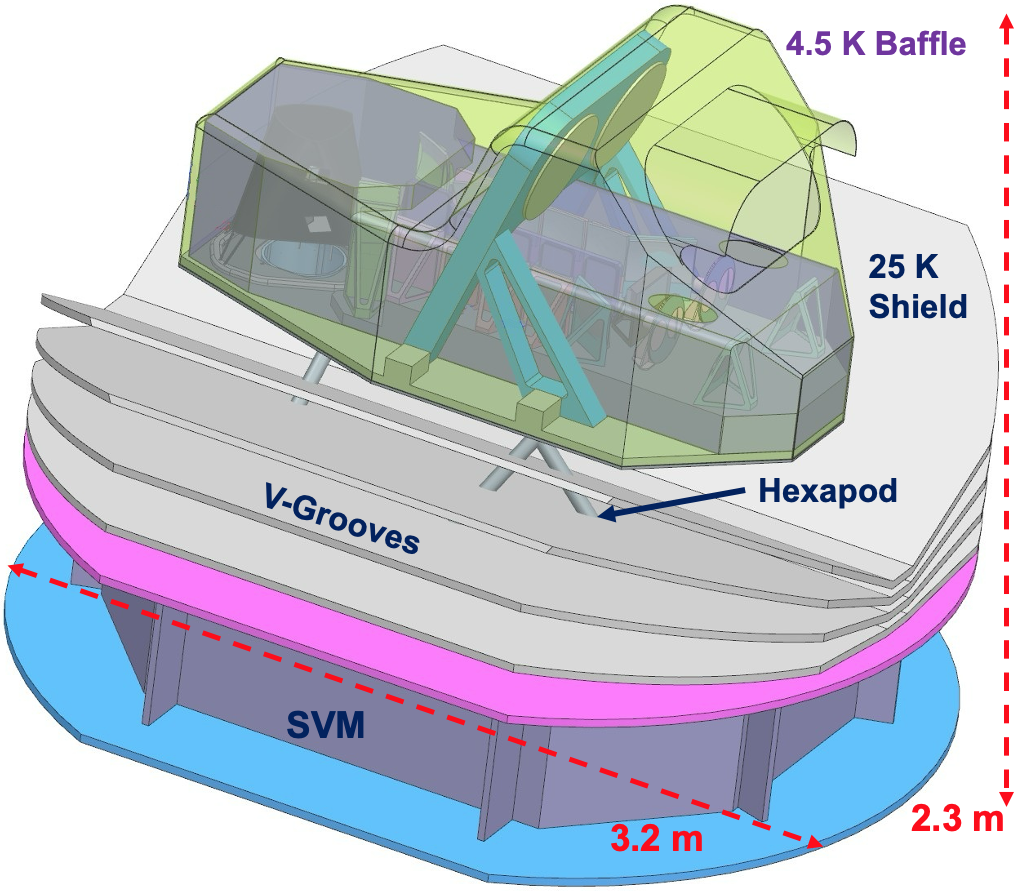}
    \includegraphics[width=0.54\linewidth]{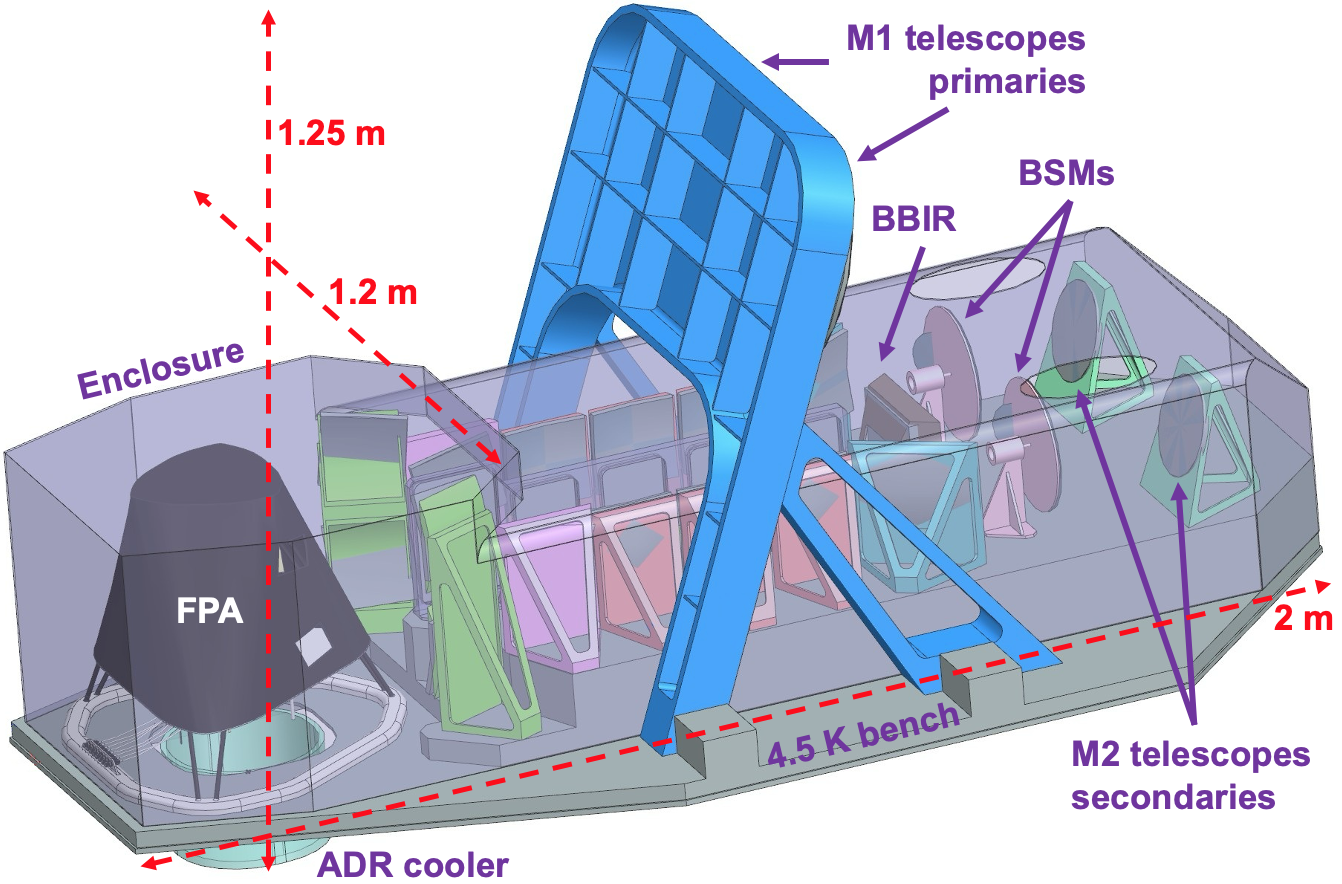}
    \caption{\textit{Left}: Overall view of \ourmission on a spacecraft benchmarked and upscaled from ARIEL's. \textit{Right}: View of \ourmission's instrument with is main elements identified in purple.}
    \label{fig:platform}
\end{figure*}

Similar to \Planck, the \ourmission instrument shown in Fig.~\ref{fig:platform} (right panel) is cooled thanks to a multi-staged cooling chain, with the detectors cooled to sub-K temperatures (Sect. \ref{sec:cooling_chain} and \cite{Sauvage2026-termal} for the detailed thermal architecture). 
\\
The \ourmission instrument is mounted on an aluminium-based 4.5\,K bench and is surrounded by a 4.5\,K baffle. It consists of a polarised Martin-Puplett FTS (Sect.~\ref{sec:fts}) with two input ports that can be fed either from a telescope pointing at the sky or from the \BBIR, which is thermally isolated from the bench so that its temperature can be actively and independently controlled (Sect.~\ref{sec:BBIR}).

The two telescopes, one for each arm of the FTS, are off-axis dual-mirror Cassegrain configurations that obey the Mizuguchi-Dragone condition to avoid cross polarisation before the first polarising element of the \FTS. The primary mirrors (M1) have a physical diameter of about 420\,mm. The projected aperture diameter gives a $\simeq1.6^{\circ}$ equivalent Gaussian beamwidth on the sky. The primary mirrors are coupled to $\sim$200\,mm diameter secondary mirrors (M2). Given the needed thermal stability, aluminium is preferred for their manufacture.

Following the optical path from the telescope, a Beam Switch Mechanism (\BSM) lets the incoming sky beam pass through, towards one of the \FTS inputs. On the other hand, if the desired operation mode needs the input to look at the \BBIR, the \BSM position selects a mirror. The combination of the two \BSMs can select all four modes of observation. 
Switching the beams can be obtained, for example, by using two rotation wheels, which will not be operated continuously, positioned before the entrance to the FTS, with a hole and a mirror that allow the beam from the telescope-sky or from the \BBIR to be directed to the associated FTS input. 

Once both inputs go through each arm of the FTS, each output is directed towards a dichroic filter splitting the full spectral range into two sub-bands, a Low-Frequency (\LF) band and a High Frequency (\HF) band. This results in four output beams, each focused onto a detection unit (DU - a multimode feedhorn coupled to the detectors) forming one Focal Plane Assembly (\FPA - Sect.~\ref{sec:FPA}). The \FPA, surrounded by its own thermal shield at 1.8\,K, is located on the 4.5\,K bench, but thermally isolated through a mechanical structure, so that detectors and their feedhorns are cooled to 50\,mK. The FPA, the FTS, and all elements up to the \BSMs are surrounded by a light-tight enclosure cooled at 4.5\,K that ensures stray-light rejection. In addition, spectral filters, located at specific enclosures and feedhorn apertures, transmit frequencies below a cut-off and reflect those above, or vice versa.
They are part of the so-called quasi-optical components which also include the FTS polarisers and the dichroic filter (at the \FTS outputs), that operate in transmission at defined angles of incidence. Their design relies on dielectrically embedded mesh filters achieved using frequency selective surfaces, whose manufacture benefits from a large heritage and is based on classic photo-lithography techniques \cite[e.g., ][]{2006SPIE.6275E..0UA,2010A&A...520A..11A}. 
%
%
\subsection{FTS}
\label{sec:fts}
\begin{figure}[!ht]
    \centering
    \includegraphics[width=1.1\linewidth]{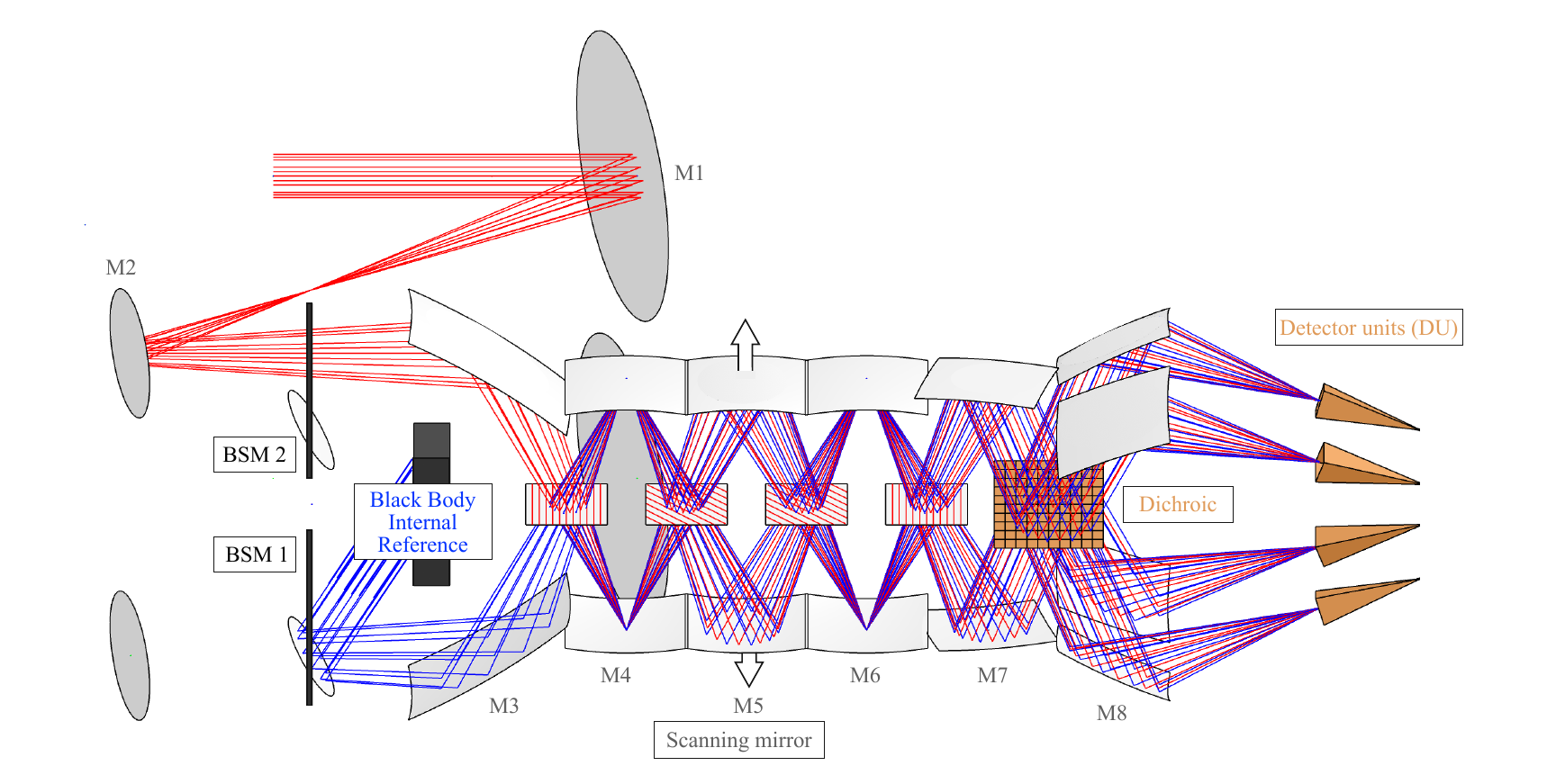}
    \caption{Overall optics ray-tracing: from the telescope (M1 and M2) the beam goes through the FTS arm where each mirror (M3 to M7) images the previous one to the next. The different polarisations from each input is mixed and separated by the four polarisers. The dichroic filter splits the spectral band into a high- and a low-frequency band resulting in 4 outputs which are collimated on the feedhorns through M8 mirrors. Figure taken from \cite{Loquet2026-optical}.}
    \label{fig:FTS}
\end{figure}

The spectrometer is a wide-band \FTS in a Martin--Puplett configuration \citep{Martin1970_martin_puplett} providing a suitable solution already implemented in \COBEF.
The optical layout of the polarising FTS is illustrated in Fig.~\ref{fig:FTS}.  A series of mirrors, M3-M7, image the previous mirror to the next and propagate the optical beams through four polarising wire grids. The schematic diagram in Fig.~\ref{fig:concept} shows how the polarisation of the beam is split by Polariser A, mixed, and then recombined by Polarisers B and C, and split again at Polariser D. The outer polarisers (A and D) transmit vertical polarisation, while the inner pair (B and C) is oriented at 45$^\circ$ relative to them. Two M5 mirrors (one in each FTS arm) are moved from their nominal positions by the FTS mechanism (FTSM) \cite[see][for examples of FTSM designs]{Cournoyer2023,2025SPIE13623E..0IK}. This introduces an extra path length into one arm and reduces the path length through the other arm, resulting in an optical phase delay. The FTS spectral resolution is tuned by adapting the moving mirror stroke so that it reaches the required resolution of $\Delta\nu=15$\,GHz up to $\nu_{max}=2$\,THz, with the possibility to slightly modify it to 10~GHz if needed. This coarse spectral resolution is suitable for the measurement of the continuum and broad spectral atomic and molecular lines.
%
\subsection{Blackbody internal reference - \BBIR}
\label{sec:BBIR}
The \BBIR provides an absolute sky brightness reference for scientific measurements and a fully characterised, temperature-controllable source for evaluating systematic effects along the instrument's optical path. To perform these tasks, it must represent the CMB emission from a BB at $T_{\rm CMB}$ and a high-frequency dust component, as it passes through the Galaxy, provided by a higher-temperature absorber at about 20\,K. 

\begin{figure}[th]
\centering
\includegraphics[width=0.8\textwidth]{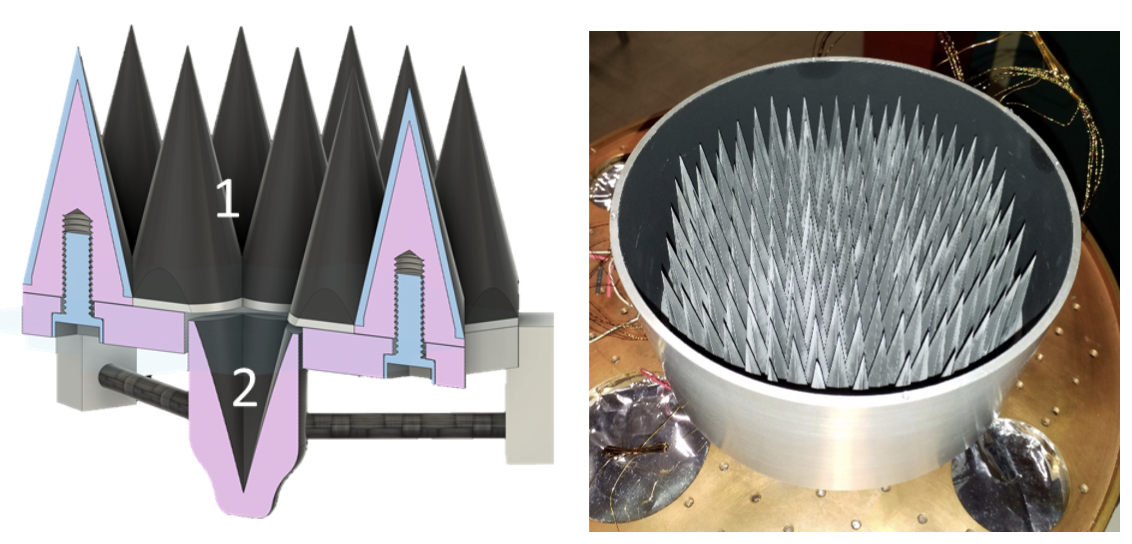}
    \caption{\textit{Left}: Schematic view of the absorbing target concept design (see text for definition of regions 1 and 2). \textit{Right}: Example of a 4\,K cold load for the TMS project \citep{TMS-BBIR}.}
    \label{fig:BBref} 
\end{figure}

The \BBIR design is derived from previous studies and experiments \citep{Fixsen2011_ARCADE2,Kogut2020_PIXIECalibration,Simonetto2021_MWIReflectivity,Valenziano2009_LFI4KRLU}
It is conceived to fulfill the dual task mentioned above; hence it is based on two distinct regions that effectively act as independent BBs, controlled at $T_{\rm CMB}$ and $\sim$20\,K respectively, each characterised by geometry and emission characteristics that meet different scientific requirements. In the baseline configuration, the  $T_{\rm CMB}$ region (Fig.~\ref{fig:BBref}, region 1) surrounds the 20K region (Fig.~\ref{fig:BBref}, region 2), which occupies a limited area of the entire \BBIR and is thermally decoupled from region 1. In addition, a baffle, actively controlled around $T_{\rm CMB}$, which consists of a metal shield that can be blackened to improve its performance, plays the dual role of integrating cavity and element capable of shielding external spillover radiation. 
The primary absorbing region (Fig.~\ref{fig:BBref}, region 1) has a temperature that can be actively controlled over the range $\sim$2.5--2.9\,K, exploiting the 1.8~K multi-stage ADR cooling power~\citep{Sauvage2026-termal}. It is implemented as an ensemble of tapered absorbing elements (e.g., pyramidal or conical structures, single‑ or multi‑layered) whose tapering is optimised to maximise emissivity and minimise reflections. Heat flow from the instrument subsystems at 4.5 K to the colder \BBIR drives thermal gradients within the absorber. Our preliminary study shows that the expected level of \BBIR gradients is a few $\mu$K. The resulting bias in $\mu$-type distortions from gradients of this order, once we fit the spectral difference between the \BBIR and the sky, is quite negligible and will not affect \ourmission limits for $\mu$ provided the instrument is cooled at 4.5\,K or colder. \\
Reflections from the \BBIR terminate elsewhere within the instrument. The reflected signal depends on both the \BBIR reflection coefficient and the temperature distribution within the 4.5~K instrument. Thermometers on the instrument walls and optical components monitor temperatures (Sect. \ref{sec:calib}) to enable the subtraction of spurious signals. 
Our preliminary study shows that to achieve unbiased detection of the expected $\mu$-type distortion, 
the \BBIR power reflection must be lower than -65~dB as a minimum requirement, with a target requirement -70~dB, and the instrument temperature gradients of order 0.1~K. Region 2, on the other hand, can obey a more relaxed emissivity requirement (sum of total losses less than -30 dB as a minimum requirement).

\subsection{Focal Plane Assembly}
\label{sec:FPA}
\begin{figure*}[h]
\includegraphics[width=0.46\textwidth]{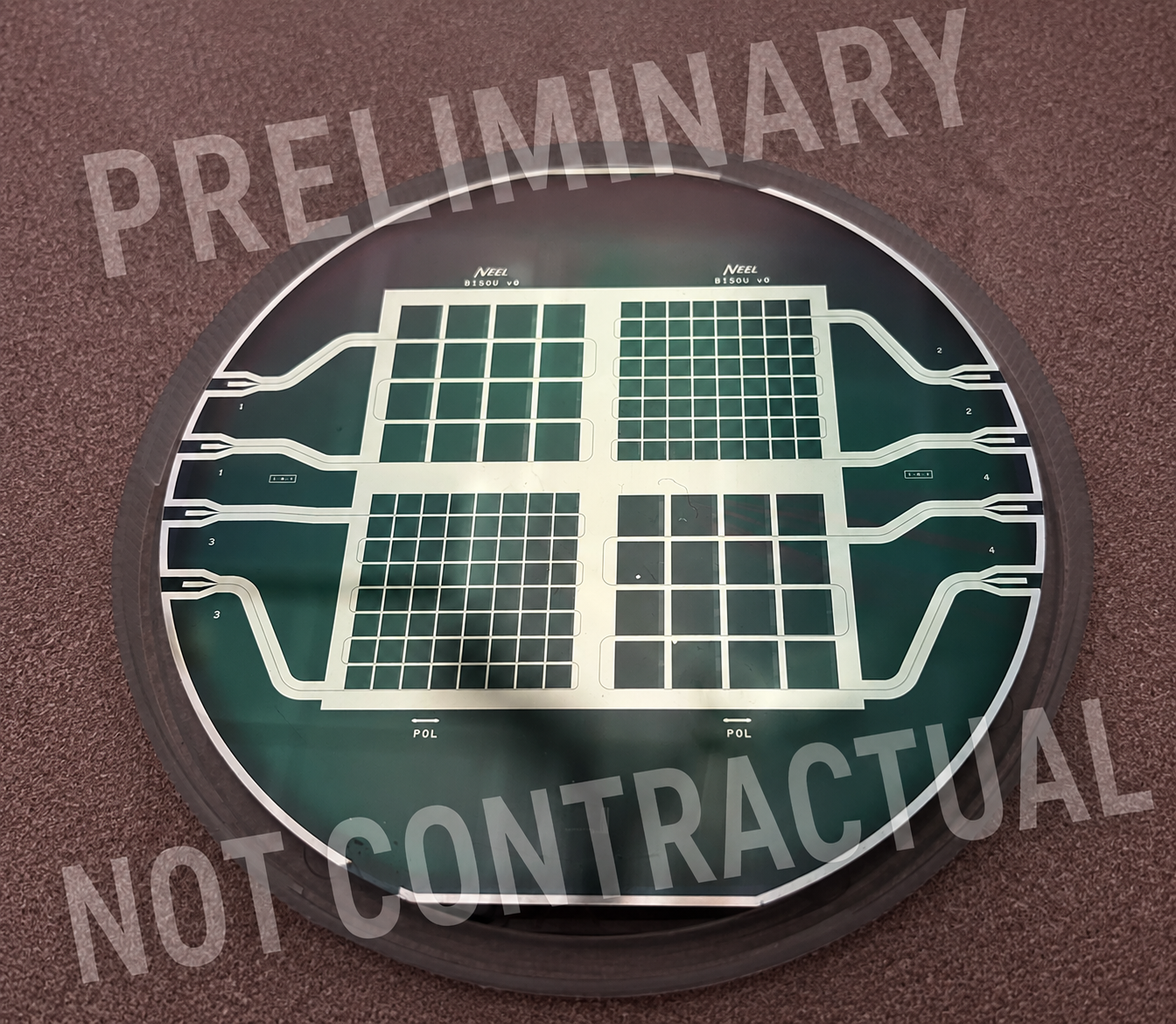}\includegraphics[width=0.5\textwidth]{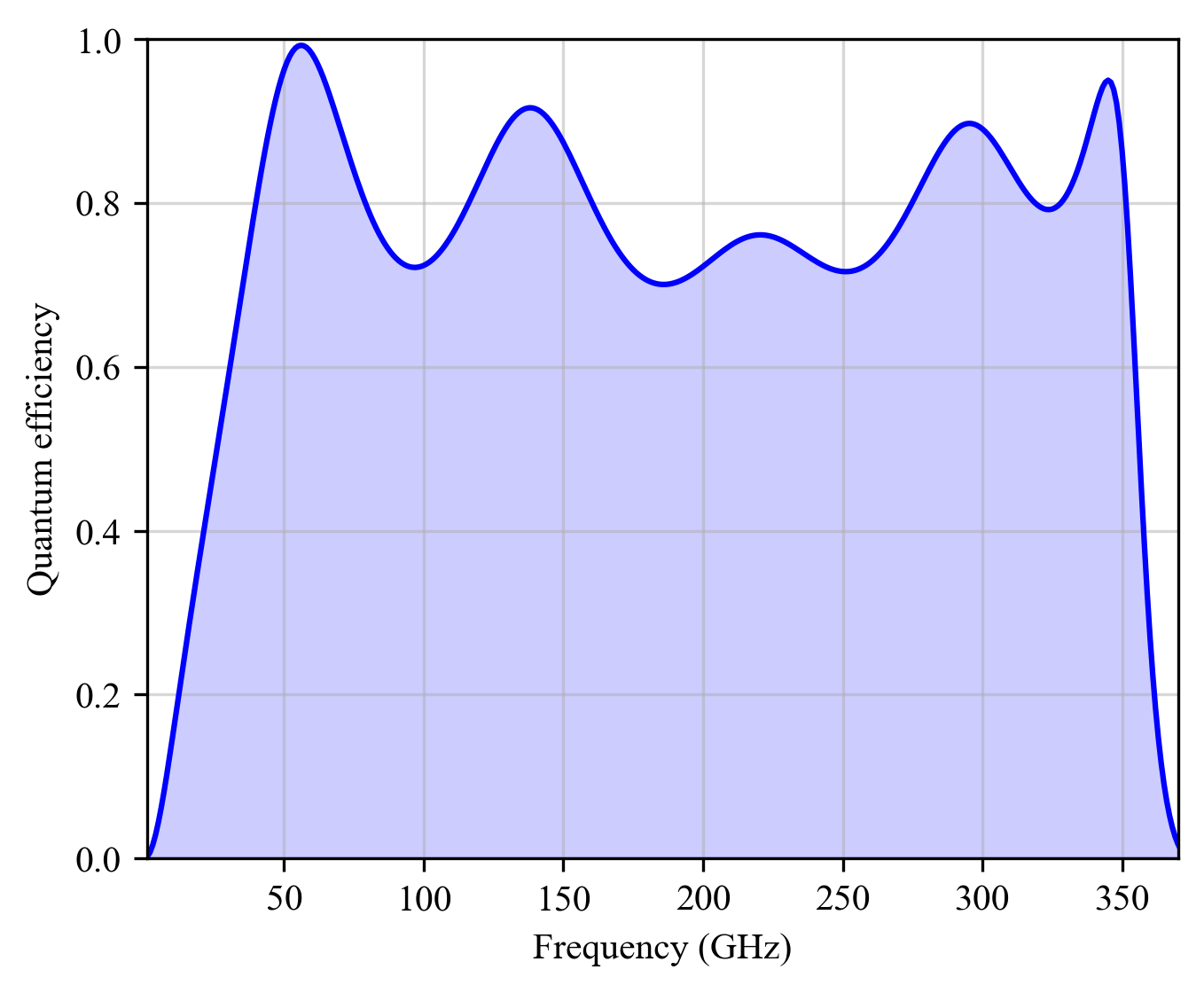}
    \caption{\textit{Left}:  LEKID prototype structures for the \ourmission LF band. The testing wafer contains four \ourmission sub-arrays, two with 16 and two with 36 detectors. It integrates on the back (not shown) the multi-layer impedance matching stack ensuring the needed wide band frequency response. \textit{Right}: The simulated quantum efficiency of LEKIDs in the low frequency spectral range $\leq$360 GHz.}
    \label{fig:LF-detec} 
\end{figure*}

The Focal Plane Assembly (\FPA) hosts four units which couple sub-arrays of detectors to feedhorns, two are Low-Frequency Detector Units (LFDU) covering the spectral band 50 to 300\,GHz and two are High-Frequency Detector Units (HFDU) covering the frequency range $300-2000$\,GHz, all cooled at 50\,mK. The signals from the individual detectors in each sub-array are summed to produce a single output IFG.
\\
The baseline individual detectors of the sub-arrays for \ourmission are Kinetic Inductance Detectors (KIDs) relying on superconducting resonators whose resonant frequencies depend on the density of Cooper pairs and quasiparticles \citep{Day2003LEKID}. They combine intrinsic frequency-domain multiplexing capability, high sensitivity (Noise Equivalent Power reaching NEP$_{\rm det}\simeq 10^{-20}$\,W\,Hz$^{-0.5}$ \citep{Baselmans2022}), fast response (quasiparticle lifetimes of hundreds $\mu$s), high dynamic range, photon-noise-limited operation, and robustness against environmental disturbances \citep{Karatsu2016,MonfardiniSPIE2016}. The detectors are coupled to multimoded feedhorns whose exit waveguide sets the type and number of modes, $n$, which can propagate from lowest to highest frequency while keeping an almost constant high throughput $A\Omega = n\lambda^2$ of about 3\,cm$^2$.sr, both for the LFDU and HFDU. The HFDUs will consist of distributed superconducting resonator KIDs as described in \cite{Dabironezare2025}, and adopted for the PRIMA and POEMM missions \citep{Glenn2025PRIMA}. The LFDUs will consist of a segmented and fully multiplexed Lumped Element KID (LEKID) device following those successfully deployed in ground-based \cite[e.g.,][]{NIKA2011,NIKA22018,Catalano2020,Concerto2024} and suborbital \cite[e.g., ][]{Paiella2020}) experiments operating at CMB frequencies. A prototype LEKID structure designed for \ourmission  with a multi-layer impedance matching stack, ensuring the requirement of wide band frequency response, is shown in Fig.~\ref{fig:LF-detec}.
\\
Based on heritage from \textit{Planck}-HFI~\citep{PlanckEarlyResultsII2011} and Closed-Cycle Dilution Refrigerator - Structural and Thermal Model ~\citep{Sauvage2025_StructuralThermalCCDR}, the FPA structure consists of an isostatic hexapod structure, thermally decoupled from the 4.5\,K bench, surrounded by a baffle that provides the magnetic shield and the thermal, mechanical, and electrical interfaces with the rest of the instrument.
%
\subsection{Cooling chain}
\label{sec:cooling_chain}
As in the case of the successful \textit{Planck} mission~\citep{Planck_thermal2011}, the cooling chain of \ourmission follows a staged thermal architecture, with each successive cooling stage progressively reducing the thermal load intercepted by the following stage \citep[see details in][]{Sauvage2026-termal}. It includes adaptations drawing from the ARIEL mission design~\citep{ARIEL_thermal2022} and the Next Generation Cryogenic IR Telescope (NG-CryoIRTel) study~\citep{NG_CryoIRTel2014,Saijo2021_SPICA_CryogenicCooling}.
\\
The coldest stage is at a temperature of 50\,mK, needed for the detection chain. It is achieved thanks to a multi-stage Adiabatic Demagnetisation Refrigerator (ADR) system, providing continuous cooling power at 1.8\,K and 0.350\,K, together with a single-shot cooling power at 50\,mK (with 80\% duty cycle over 35~hours)~\citep{ADR_SPIE_2026}. This technological solution, studied for SPICA/SAFARI \citep{DUBAND2014} and LiteBIRD \citep{2020JLTP..199..730D} projects, was also chosen for the NewAthena/X-IFU instrument \citep{Prouve2020}. The 4.5\,K thermal stage, needed for the rest of the instrument, will be provided by the 4\,K cooler under ESA responsibility, which will also provide cooling power at an intermediate stage of 25\,K. The cooling down from ambient Service Module (SVM) temperature to 50~K is enabled by three V-groove passive radiators arranged in cascade (Fig.\,\ref{fig:platform}) so that they progressively intercept and re-radiate heat toward deep space. 

%
\subsection{Temperature monitoring and control system}
The temperature monitoring and control system (TMCS) will be responsible for acquiring temperature telemetry from the relevant instrument components. These include the detector units as well as the main optical and structural elements. In addition, the TMCS will regulate the temperature of the \BBIR and maintain it with the required stability. The \BBIR reference temperature will lie in the range 2.5--2.9\,K and must be controlled with high absolute accuracy (see next section).
As a reference, thick-film ruthenium oxide resistors are expected to provide a temperature measurement precision of approximately 0.1\,mK in 1\,s \citep{kogut/fixsen:2020}. 
The complete design of the temperature readout and control system builds upon the heritage of the \textit{Planck} mission, while also benefiting from ongoing developments for future CMB missions such as \textit{LiteBIRD}. The resulting housekeeping data provided by the TMCS will be used together with the scientific measurements to reconstruct the IFG and to mitigate systematic effects.
%
\section{Calibration} \label{sec:calib}
The ground calibration strategy of \ourmission will build on standard procedures and heritage of previous missions such as \Planck. 
The payload will go through sequences of validation and calibration, from component to spacecraft level, supported by detailed simulations to establish a complete instrument model allowing the instrument systematics to be assessed. The instrument models will be completed by the science and the housekeeping data acquired in flight, as seen below.
\\

Flight data will be used for calibration from raw detector telemetry to sky surface brightness \cite[see][for a reference study for the in-flight calibration methodology]{kogut/fixsen:2020}. 
\\
In the two science modes (B01 and B10 in Fig. \ref{fig:concept}), the uncalibrated IFGs from a given sky-pixel encode the spectral difference between the sky and the \BBIR. The sky signal is cancelled in the comparison of data between B01 and B10, leaving the differential spectrum from the \BBIR whose temperature is at a given set-point. This cancellation can be performed for a series of distinct set-points, i.e. \BBIR temperatures. The differential
Planck spectrum $\partial B(T,\nu) / \partial T$ observed at different \BBIR set-points bracketing the CMB monopole $T_0$, i.e. between 2.5 and 2.9\,K, provides a reference signal for the intensity scale in each frequency bin, while the shift in peak frequency (Wien displacement law) provides a reference for the frequency scale \citep{kogut/fixsen:2020}. Additionally, observations of bright Galactic emission lines ({\sc C[ii], N[ii]}, CO) can provide a secondary reference for the frequency scale. Interpolation of the sky/\BBIR difference with the \BBIR at different temperatures in the range 2.5--2.9\,K allows the determination of $T_0$, and hence any spectral distortions, to precision of a few nK \citep{kogut/fixsen:2020}. Thermometers embedded within the \BBIR, with 100\,$\mu$K\,$\sqrt{s}$ readout noise, monitor the temperature across the \BBIR. Uncertainty in the absolute thermometry scale (from ground calibration) yields a $\Delta T_0 \approx 100\,\mu$K uncertainty in the absolute monopole temperature, but meets the needs of the spectral distortion measurements. Future developments will incorporate absolute temperature sensors that will either replace the resistive sensors  or provide an absolute reference for their calibration.
\\
Flight data will also be used to measure instrument parameters and use them to monitor and control the systematic effects. Thermometers will measure the temperature of key elements in the optical path (mirrors, filters, etc.), as well as the instrument enclosure and baffle throughout the mission to monitor the instrumental emission. The flight data taken with the \BBIR at different temperatures can be used to estimate the corrections for possible reflections, to model them, and to subtract them from the sky signal. Stray-light corrections can be obtained from the BSM mode B00, with both beams looking at the \BBIR  (see Fig. \ref{fig:concept}). Comparison of IFGs with forward vs. backward FTSM scanning provides a measurement of instrumental time constants, including the detector and readout electronics as well as thermal control of the \BBIR and optical elements. \ourmission samples a double-sided IFG; comparison of the real and imaginary parts of the resulting Fourier transform determines the position of zero phase difference in the FTSM scan. 
\\
During the next stages of \ourmission, the preliminary instrument model developed for the forecast and concept optmisation will be detailed to include the relevant parameters for the characterisation of the instrument response and to model the IFG measurement. This will be continuously updated with improved models/parameters, based on ground calibration, as well as information from housekeeping collected throughout the mission. The instrument model will allow us to test that the calibration procedure is sufficiently sensitive, accurate, and robust against systematic effects (like temperature differences between telescopes, optical elements before the input polariser).
\\
Finally, observations of planets will be used to map the beam patterns in-flight.

\section{The FOSSIL mission}\label{sec:mission}
The overall volume required for \ourmission, about $2.4\,\mathrm{m}\times1.2\,\mathrm{m}\times1.1\,\mathrm{m}$, is driven by the size of the bench and the 4.5\,K baffle that surrounds the whole instrument. This requires a heritage \Planck SVM or an upscaled ARIEL SVM that reaches a dry mass of approximately 1500\,kg compatible with a launch by an Ariane 6.2 launch vehicle \citep{A6_Users_Manual}. Both SVM options provide the required total power budget of the payload and can handle its data transmission rate, of 7 Mbit/s after an onboard averaging and compression factor of five, which can be kept within an X-band system using the entire 10\,MHz frequency bandwidth, or alternatively can be transmitted using the Ka band downlink.
\\
A small amplitude Lissajous orbit around L2 (Sun spacecraft Earth angle $\sim 15^{\circ}$) with a platform oriented towards the Sun is chosen, with a spin axis quasi-parallel to the Sun-Earth direction ($\pm$ 5$^{\circ}$) and a nominal Solar Aspect Angle (SAA) of approximately $\pm$10$^{\circ}$. The solar panels on the Sun-facing surface shield the payload from the main heat source while allowing the payload V-grooves to passively cool by dumping excess heat, similar to that achieved by \Planck \citep{Bersanelli_LFI_design}. 

\ourmission is a survey instrument with four observation modes that will be operated during the entire planned 4-year mission. To achieve full coverage of the sky in six months, the satellite will be in a spinning configuration similar to \Planck but with a slower spin rate ($\sim 2$\,rph). The spin axis being quasi-parallel with the Sun-Earth axis and the payload Line of Sight (LoS) in an almost normal angle to the spin axis, allows for great circles in the sky similar to, but significantly wider than, those of \Planck. 

\begin{figure}[th]
    \centering
    \includegraphics[width=\linewidth]{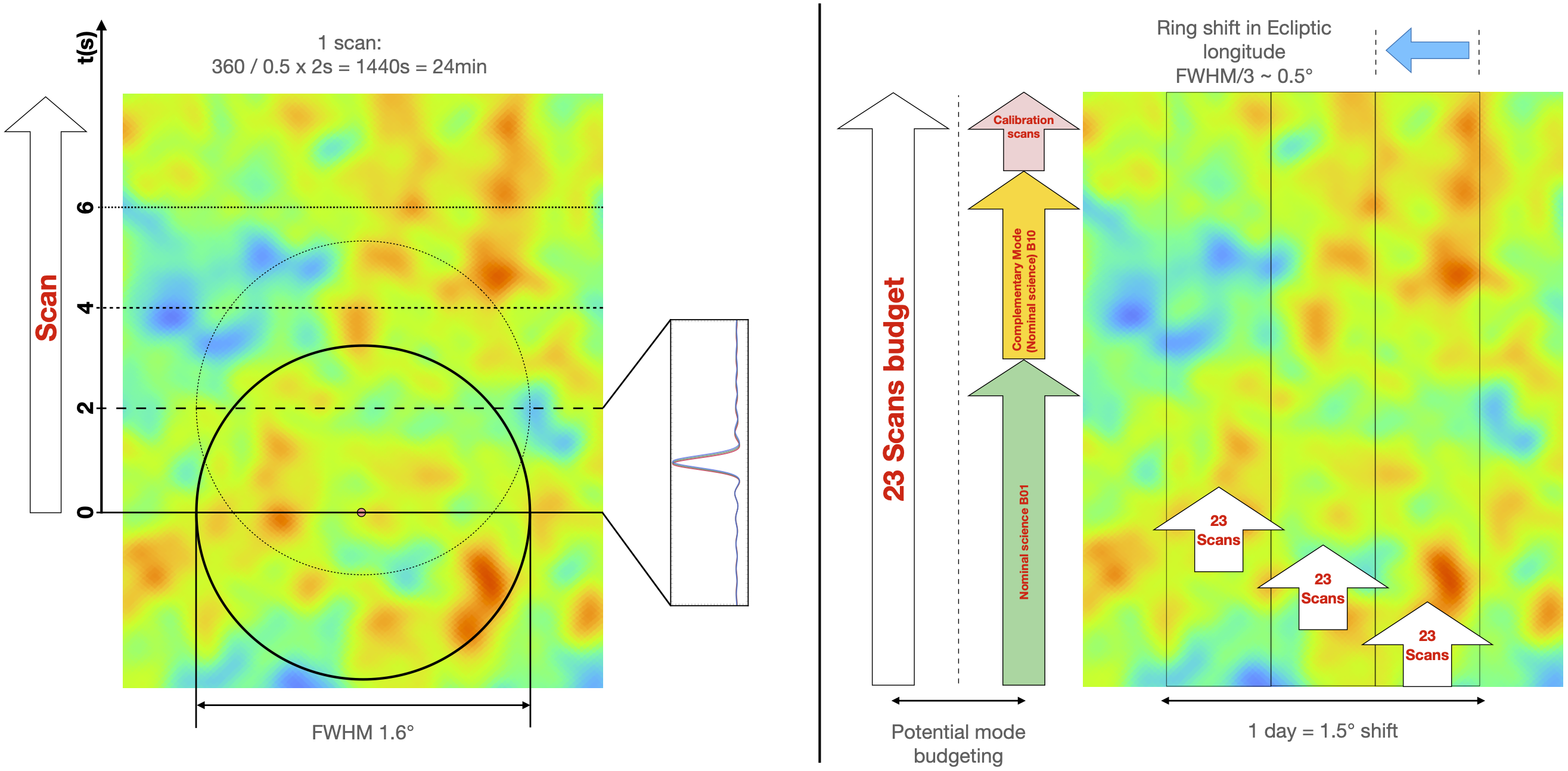}
    \caption{Principle of the scanning strategy. \textit{Left}: \ourmission beam size and timing of IFG acquisition. \textit{Right}: A single ring scan is performed in $\sim 24$ min and repeated 23 times before moving in ecliptic longitude by $\sim 0.5^{\circ}$, shift shown on the far right.}
    \label{fig:scan}
\end{figure}

The scanning strategy (see Fig.~\ref{fig:scan} for a schematic view) is designed to provide high-cadence spectral sampling while meeting spatial Nyquist requirements of the $1.6^{\circ}$ baseline beam. The fundamental unit of observation is the acquisition of a single IFG over a duration of $\tau_{\mathrm{IFG}} \sim 2\,\mathrm{s}$. With a nominal telescope scan speed of $0.25^{\circ}/\mathrm{s}$, each IFG corresponds to a spatial displacement of $0.5^{\circ}$ in the sky. Given a beam of $\theta_{\mathrm{beam}} = 1.6^{\circ}$, this yields roughly $3.2$ samples per beam, satisfying the supra Nyquist criterion. A scan comprises almost a full maximum circle in the sky. Taking into account the $0.5^{\circ}$ sampling, a single scan is completed in $24\,\mathrm{min}$ and consists of 720 IFGs. Scans are grouped into rings, with a single ring consisting of 23 successive scans, totalling approximately $9.2\,\text{hours}$ of integration. In order to achieve optimal coverage of the sky, successive rings are shifted by about $0.5^{\circ}$ so that three consecutive rings total approximately $28\,\mathrm{hours}$ of observation time in the 35 hour-cycle of the ADR cooler. 
%
\section{Data analysis}\label{sec:data}
\begin{figure*}[th]
    \centering
    \includegraphics[width=\linewidth]{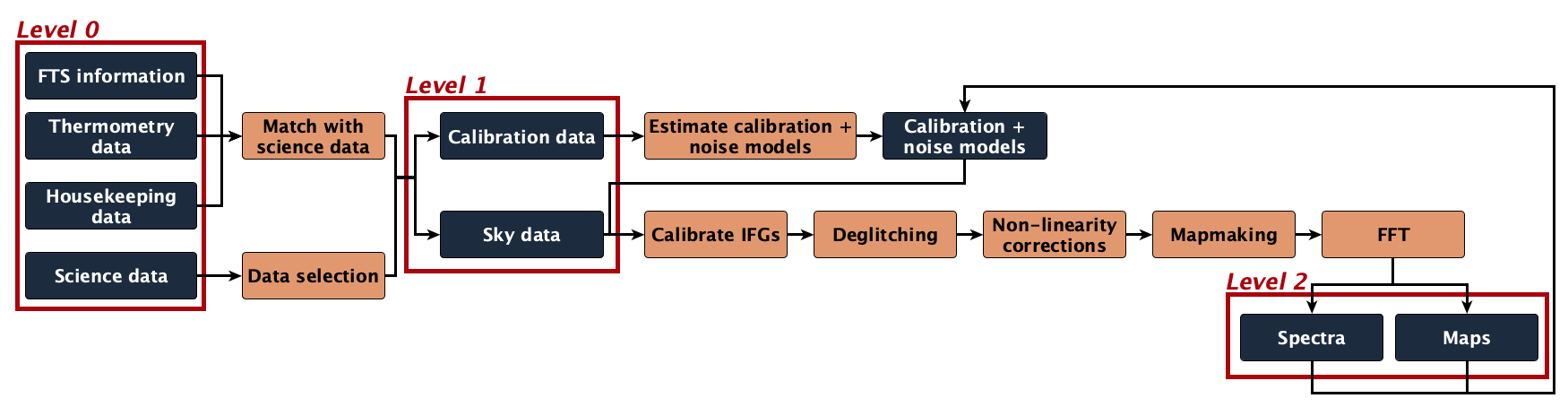}
    \caption{Schematic view of the preliminary data analysis pipeline. The blue boxes represent intermediary or final data products, whereas the orange ones represent essential pipeline steps.}
    \label{fig:data-pipeline}
\end{figure*}

\ourmission is a survey-type mission with ground operations, involving shared responsibility between ESA and the consortium, which include both satellite and instrument operations, as well as data processing and scientific analysis. A preliminary sketch of the \ourmission data processing pipeline (Fig. \ref{fig:data-pipeline}) is inspired by those of legacy FTS experiments, such as \COBEF \citep{firasexsupp}, and updated with the more recent CONCERTO pipeline \citep{Concerto2024}. The preliminary pipeline of \ourmission shows the end-to-end flow, from raw data to science data analysis. Data reduction and analysis include the production of IFG timelines, their cleaning (e.g., removing glitches) and calibration, and the production of spectra.

\ourmission will produce three IFGs per beam during rotation in great circles which will correspond each to a spectrum of all sources in the integrated beam. These {\it spaxels} (spectral pixels) in the sky will be created as a result of great circle (ring) redundancy to increase S/N and remove any low frequency temporal drifts of any instrumental effects.
In either of the two calibration modes (B00 or B11), the differential nature of the FTS provides only the spectral content of the two FTS arms above the detector loading offset \citep[see e.g., ][]{naess2019}. This will provide unprecedented knowledge on calibration details (including noise sources, systematics, and emissions from instrument elements) usually only available during ground level tests.
In the science modes (B01 or B10), the recorded spectra will be the difference between the sky and the temperature of the \BBIR for each spaxel, from 50 to 2000\,GHz. The science analysis, including component separation and parameter determination, can then be performed with \ourmission products alone and in combination with external datasets.
\\
Throughout the duration of \ourmission, simulations will be essential at each step from pipeline development, algorithm validation, to systematic error characterisation and cosmological inference. They will consist of both end-to-end simulations combining sky and instrument that reproduce the full observational process and Monte-Carlo campaigns used to assess and propagate uncertainties in map-making, component separation, and eventually cosmological interpretation.

\section{Conclusion}
Arising from physical mechanisms at all scales and epochs, CMB spectral distortions represent a new observational window on our Universe. They allow us to probe the standard cosmological model and to investigate both new physics and baryonic effects throughout the thermal history of the Universe.

In the last two decades, several initiatives have been attempted to propose a space mission dedicated to the detection of CMB spectral distortions. Among them PIXIE, PRISM, PRISTINE, CMB-BHARAT, etc. were proposed to NASA, ESA, or ISRO. However, since the 1990s and the success of the COBE satellite,  
no space mission or ground-based telescope targetting CMB spectral distortions as their main science goal has been developed. The ESA call for the 8th Medium-class mission is a unique opportunity for a consortium of specialists in instrumentation, theory, and data analysis to launch a CMB spectroscopy mission. 

This mission, \ourmission, is designed to be at the forefront of cosmology in the 2040s. At its expected three orders of magnitude improvement over the seminal measurement of the CMB spectrum, the latter cannot remain a perfect blackbody and \ourmission offers a unique opportunity to mine exceptionally rich science information.

By measuring $\mu$ and $y$ distortions at unprecedented sensitivity, it will enable us to put stringent constraints on the dissipation of primordial perturbations by Silk damping, on early and late time signatures of DM production, annihilation, decay, and interaction mechanisms, and on the thermal energy injected during the large-scale structure evolution. \ourmission will thus address the following key questions: \textit{What governs the physics of inflation?} \textit{What is the precise nature of dark matter?} \textit{How did the growth of structures proceed?} In addition, the legacy value of \ourmission with 130 frequency bands between 50\,GHz and 2\,THz will allow scores of synergetic analyses with all major ground and space facilities.

Spectral distortions are a game-changing  observable that will provide us with a novel tracer of the thermal state of the Universe throughout its evolution. This will be possible with a CMB spectroscopic survey like \ourmission that can  achieve a major step forward in observational cosmology and potentially discover new physics beyond our standard models, and hence recast our understanding of the Universe.

\appendix

\section{Author affiliations}
\label{app:affiliations}

\noindent
$^{1}$ Université Paris-Saclay, CNRS, Institut d'Astrophysique Spatiale,
91405 Orsay, France

\smallskip
\noindent
$^{2}$ Université Paul Sabatier, CNRS, CNES, Institut de Recherche en
Astrophysique et Planétologie (IRAP), Toulouse, France

\smallskip
\noindent
$^{3}$ Aix Marseille Univ, CNRS, CNES, LAM (Laboratoire d'Astrophysique
de Marseille), Marseille, France

\smallskip
\noindent
$^{4}$ Department of Astronomy, University of Geneva, Chemin d’Ecogia 16,
1290 Versoix, Switzerland

\smallskip
\noindent
$^{5}$ Jodrell Bank Centre for Astrophysics, Department of Physics and
Astronomy, University of Manchester, Oxford Road, Manchester M13 9PL, U.K.

\smallskip
\noindent
$^{6}$ INAF - Osservatorio di Astrofisica e Scienza dello Spazio (OAS) di
Bologna, via P. Gobetti 93/3, I-40129 Bologna, Italy

\smallskip
\noindent
$^{7}$ Center for Theoretical Physics - A Leinweber Institute,
Massachusetts Institute of Technology, Cambridge, MA, USA

\smallskip
\noindent
$^{8}$ Dipartimento di Fisica, Sapienza Università di Roma,
Piazzale Aldo Moro 5, I-00185 Roma, Italy

\smallskip
\noindent
$^{9}$ INFN Sezione di Roma, Piazzale Aldo Moro 5, I-00185 Roma, Italy

\smallskip
\noindent
$^{10}$ Instituto de Astrofísica de Canarias (IAC),
Departamento de Astrof\'{\i}sica, Universidad de La Laguna,
E-38206 La Laguna, Tenerife, Spain

\smallskip
\noindent
$^{11}$ University of Groningen, Groningen, NL

\smallskip
\noindent
$^{12}$ INFN, Sezione di Bologna, via Irnerio 46, 40126 Bologna, Italy

\smallskip
\noindent
$^{13}$ University of Iceland, Reykjavík, IS

\smallskip
\noindent
$^{14}$ Observational Cosmology Laboratory, Goddard Space Flight Center,
Greenbelt, MD 20771, USA

\smallskip
\noindent
$^{15}$ Centre national d’études spatiales (CNES), France

\smallskip
\noindent
$^{16}$ Univ. Grenoble Alpes, CEA, IRIG, DSBT, 38000 Grenoble, France

\smallskip
\noindent
$^{17}$ Institute of Theoretical Astrophysics, University of Oslo,
Sem Sælands vei 13, 0371 Oslo, Norway

\smallskip
\noindent
$^{18}$ Institut Néel, CNRS and Univ Grenoble Alpes,
25 rue des Martyrs, 38042 Grenoble, France

\smallskip
\noindent
$^{19}$ Maynooth University, Maynooth, Ireland

\smallskip
\noindent
$^{20}$ Dipartimento di Fisica e Scienze della Terra,
Universit\`a degli Studi di Ferrara, via Saragat 1,
I-44122 Ferrara, Italy

\smallskip
\noindent
$^{21}$ Istituto Nazionale di Fisica Nucleare, Sezione di Ferrara,
via Saragat 1, I-44122 Ferrara, Italy

\smallskip
\noindent
$^{22}$ Instituto de Física de Cantabria (CSIC-UC),
Avda. de los Castros s/n, 39005 Santander, Spain

\smallskip
\noindent
$^{23}$ Université Paris-Saclay, Université Paris Cité, CEA, CNRS, AIM,
91191 Gif-sur-Yvette, France

\smallskip
\noindent
$^{24}$ Physics \& Astronomy Dept., University College London (UCL),
WC1E 6BT London, UK

\smallskip
\noindent
$^{25}$ Institute for Space Imaging Science, University of Lethbridge,
Lethbridge, Canada

\smallskip
\noindent
$^{26}$ INFN - Sezione di Pisa, Edificio C,
L.go Bruno Pontecorvo 3, 56127 Pisa, Italy

\smallskip
\noindent
$^{27}$ Mullard Space Science Laboratory (MSSL), UCL, Dorking, UK

\smallskip
\noindent
$^{28}$ Physics Department, Boston University,
Boston, MA 02215, USA

\smallskip
\noindent
$^{29}$ INAF-IAPS, Via del Fosso del Cavaliere 100,
I-00133 Rome, Italy

\smallskip
\noindent
$^{30}$ Institut d'Astrophysique de Paris (IAP), CNRS,
Sorbonne Université, Paris, FR

\smallskip
\noindent
$^{31}$ Theoretical Physics Department, CERN,
1211 Geneva 23, Switzerland

\smallskip
\noindent
$^{32}$ INAF, Istituto di Radioastronomia,
Via Piero Gobetti 101, 40129 Bologna, Italy

\smallskip
\noindent
$^{33}$ Université Paris Cité, CNRS/IN2P3, CEA,
Laboratoire AstroParticule et Cosmologie (APC), Paris, FR

\smallskip
\noindent
$^{34}$ Université Paris-Saclay, CNRS/IN2P3,
IJCLab, 91405 Orsay, France

\smallskip
\noindent
$^{35}$ Blackett Lab., Department of Physics, Imperial College,
Prince Consort Road, London SW7 2AZ, UK

\smallskip
\noindent
$^{36}$ Department of Physics, University of Oxford,
Denys Wilkinson Building, Keble Road, Oxford OX1 3RH, UK

\smallskip
\noindent
$^{37}$ INAF, Osservatorio Astronomico di Padova,
Vicolo dell’Osservatorio 5, I-35122 Padova, Italy

\smallskip
\noindent
$^{38}$ School of Mathematical and Physical Sciences,
University of Sheffield, UK

\smallskip
\noindent
$^{39}$ Institute for Theoretical Physics, Leibniz University Hannover,
Appelstraße 2, 30167 Hannover, Germany

\smallskip
\noindent
$^{40}$ Max Planck Institute for Gravitational Physics,
Albert Einstein Institute, 30167 Hannover, Germany

\smallskip
\noindent
$^{41}$ INAF-Osservatorio Astronomico di Trieste,
Via G. B. Tiepolo 11, 34143 Trieste, Italy

\smallskip
\noindent
$^{42}$ Dipartimento di Fisica e Astronomia “Galileo Galilei”,
Università degli Studi di Padova, Padova, Italy

\smallskip
\noindent
$^{43}$ Laboratoire d’Annecy de Physique Theorique (LAPTh),
CNRS/USMB, 99 Chemin de Bellevue, Annecy, France

\smallskip
\noindent
$^{44}$ Centro de Astrof\'{\i}sica da Universidade do Porto,
Rua das Estrelas, 4150-762 Porto, Portugal

\smallskip
\noindent
$^{45}$ Instituto de Astrof\'{\i}sica e Ci\^encias do Espa\c co,
Universidade do Porto, Rua das Estrelas, 4150-762 Porto, Portugal

\smallskip
\noindent
$^{46}$ Dipartimento di Fisica e Astronomia “Augusto Righi”,
Università di Bologna, Via Piero Gobetti 93/2,
I-40129 Bologna, Italy

\smallskip
\noindent
$^{47}$ Dipartimento di Fisica, Università degli Studi di Torino,
Via P. Giuria 1, I-10125 Torino, Italy

\smallskip
\noindent
$^{48}$ INFN-Sezione di Torino,
Via P. Giuria 1, I-10125 Torino, Italy

\smallskip
\noindent
$^{49}$ INAF-Istituto Nazionale di Astrofisica,
Osservatorio Astrofisico di Torino, strada Osservatorio 20,
10025 Pino Torinese, Italy

\smallskip
\noindent
$^{50}$ Instituto de Astrof\'isica e Ci\^encias do Espa\c{c}o,
Faculdade de Ci\^encias, Universidade de Lisboa,
Campo Grande, 1749-016 Lisboa, Portugal

\smallskip
\noindent
$^{51}$ Institut de Planétologie et d'Astrophysique de Grenoble (IPAG),
CNRS, Université Grenoble Alpes, Grenoble, FR

\smallskip
\noindent
$^{52}$ Laboratoire Univers et Particules de Montpellier (LUPM),
CC 72, Place Eugène Bataillon, 34095 Montpellier - Cedex 5, France

\smallskip
\noindent
$^{53}$ Centre National de la Recherche Scientifique (CNRS), FR

\smallskip
\noindent
$^{54}$ Center for Cosmology and Particle Physics,
Department of Physics, New York University,
New York, NY 10003, USA

\smallskip
\noindent
$^{55}$ University of British Columbia, Vancouver, Canada

\smallskip
\noindent
$^{56}$ Institute for Data Processing and Electronics,
Karlsruhe Institute of Technology, Hermann-von-Helmholtz-Platz 1,
76344 Eggenstein-Leopoldshafen, Germany

\smallskip
\noindent
$^{57}$ Physics Department, University of Richmond,
138 UR Drive, Richmond, VA 23173, USA

\smallskip
\noindent
$^{58}$ Center for Data-Driven Discovery, Kavli IPMU (WPI), UTIAS,
The University of Tokyo, Kashiwa, Chiba 277-8583, Japan

\smallskip
\noindent
$^{59}$ Kavli IPMU (WPI), UTIAS, The University of Tokyo,
5-1-5 Kashiwanoha, Kashiwa, Chiba 277-8583, Japan

\smallskip
\noindent
$^{60}$ Laboratoire de Physique de l'\'Ecole Normale Sup\'erieure,
ENS, Universit\'e PSL, CNRS, Sorbonne Universit\'e,
Universit\'e Paris Cit\'e, 75005 Paris, France

\smallskip
\noindent
$^{61}$ University of Oviedo/ICTEA, Spain

\smallskip
\noindent
$^{62}$ Department of Physics, University of Crete,
71003 Heraklion, Greece

\smallskip
\noindent
$^{63}$ Institute of Astrophysics,
Foundation for Research and Technology-Hellas,
71110 Heraklion, Crete, Greece


\bibliographystyle{plainnat}
\bibliography{mergedbibliography} 

\end{document}